\documentclass[12pt]{iopart}

\usepackage{graphicx}
\usepackage{multirow}
\usepackage{url}
\usepackage{cite}

\begin{document}

\title{Magnetic skin effect in Pb(Fe$_{1/2}$Nb$_{1/2}$)O$_3$}

\author{N Giles-Donovan$^{1}$\footnote{Present address: Department of Physics, University of California, Berkeley, California 94720, USA}\footnote{Author to whom any correspondence should be addressed.}, A D Hillier$^{2}$, K Ishida$^{2,3}$, B V Hampshire$^{4,2}$, S R Giblin$^{5}$, B Roessli$^{6}$, P M Gehring$^{7}$, G Xu$^{7}$, X Li$^{8}$, H Luo$^{8}$, S Cochran$^{1}$ and C Stock$^{9}$}

\address{$^1$Centre for Medical and Industrial Ultrasonics, James Watt School of Engineering, University of Glasgow G12 8QQ, UK}
\address{$^2$ ISIS Facility, Rutherford Appleton Laboratory, Harwell, Didcot, UK}
\address{$^3$ RIKEN Nishina Center, RIKEN, Wako, Saitama, Japan}
\address{$^4$ Department of Physics, University of Warwick, Coventry CV4 7AL, UK}
\address{$^5$ School of Physics and Astronomy, Cardiff University, Cardiff CF24 3AA, UK}
\address{$^6$ Laboratory for Neutron Scattering and Imaging, Paul Scherrer Institut (PSI), 5232 Villigen PSI, Switzerland}
\address{$^7$ NIST Center for Neutron Research, National Institute of Standards and Technology, Gaithersburg, Maryland 20899-6100, USA}
\address{$^8$Shanghai Institute of Ceramics, Chinese Academy of Sciences, Shanghai, China}
\address{$^{9}$School of Physics and Astronomy, University of Edinburgh, Edinburgh EH9 3JZ, UK}

\ead{ngilesdonovan@berkeley.edu}

\begin{abstract}

Relaxor-ferroelectrics display exceptional dielectric properties resulting from the underlying random dipolar fields induced by strong chemical inhomogeneity.  An unusual structural aspect of relaxors is a skin-effect where the near-surface region in single crystals exhibit structures and critical phenomena that differ from the bulk.  Relaxors are unique in that this skin effect extends over a macroscopic lengthscale of $\sim$~100~$\mu m$ whereas usual surface layers only extend over a few unit cells (or $\sim$ nm). We present a muon spectroscopy study of Pb(Fe$_{1/2}$Nb$_{1/2}$)O$_{3}$ (PFN) which displays ferroelectric order, including many relaxor-like dielectric properties such as a frequency broadened dielectric response, and antiferromagnetism with spatially short-range polar correlations and hence can be termed a multiferroic. In terms of the magnetic behavior determined by the Fe$^{3+}$ ($S=5/2$, $L\approx0$) ions, PFN has been characterized as a unique example of a ``cluster spin-glass".  We use variable momentum muon spectroscopy to study the depth dependence of the slow magnetic relaxations in a large 1~cm$^{3}$ crystal of PFN.  Zero-field \emph{positive} muon spin relaxation is parameterized using a stretched exponential, indicative of a distribution of relaxation rates of the Fe$^{3+}$ spins.  This bandwidth of frequencies changes as a function of muon momentum, indicative of a change in the Fe$^{3+}$ relaxation rates as a function of muon implantation depth in our single crystal.  Using \emph{negative} muon elemental analysis, we find small-to-no measurable change in the Fe$^{3+}$/Nb$^{5+}$ concentration with depth implying that chemical concentration alone cannot account for the change in the relaxational dynamics. PFN displays an analogous magnetic skin effect reported to exist in the structural properties of relaxor-ferroelectrics.

\end{abstract}

\maketitle

\section{Introduction}

Lead-based relaxor-ferroelectrics have attracted great attention in recent years with exceptional dielectric properties reported~\cite{ShujunReview, Ye1, Ye2, Bokov1, Ye3, Ye4}. Prototypical relaxors based on mixed perovskites with structure A(B$_x$B$_{1-x}^{\prime}$)O$_3$ include Pb(Mg$_{1/3}$Nb$_{2/3}$)O$_3$ (PMN) and  Pb(Zn$_{1/3}$Nb$_{2/3}$)O$_3$ (PZN)~\cite{Relaxing}. These materials exhibit a random mixture of cations on the B site and this inherent disorder has been implicated as, at least partially, a reason for their unique performance~\cite{Relaxing, Bokov1, Li2016}. The phase transitions of these materials have challenged the understanding of ferroelectric transitions in the presence of disorder as relaxors characteristically display a diffuse transition with a temperature and frequency broadened dielectric response~\cite{ParkShrout, Viehland1, Viehland2, Piric1, Piric2}.  However, and the purpose of this paper, a particularly unusual aspect of relaxor materials is the presence of a distinct macroscopic near-surface skin region where the structure is different from bulk phases~\cite{Kumar20:8,Kong19:29}.  Here we demonstrate the magnetic analogue of this skin region in the multiferroic Pb(Fe$_{1/2}$Nb$_{1/2}$)O$_{3}$ (PFN).

Whereas conventionally measured in stoichiometric perovskites like PbTiO$_{3}$, a single temperature scale is associated with the onset of ferroelectric order, multiple temperature scales have been observed in relaxors and the relaxor-like PFN. A high temperature scale, defined by the Burn's temperature~\cite{Burns1,Burns2} where deviations in the index of refraction are observed and a zone center transverse optical phonon mode softens in energy~\cite{Vakhrushev,Nab_soft_mode,Gehring_soft_mode,Hlinka_soft_mode,Gehring_soft_mode_2}, defines the onset of a root-mean-squared electric polarization caused by spatially localized and dynamic correlations~\cite{Burns1, Burns2, Cross}. These so-called polar nanoregions have been directly observed in neutron and X-ray diffuse measurements and are characterized by an anisotropic momentum broadened~\cite{StockPNR, Hirota, Welberry} scattering cross sections indicative of spatially short-range correlations. The second, lower, temperature scale corresponds to when long-range ferroelectric correlations become purely static and can be stabilized under an applied electric field~\cite{Ye5, StockEField}. This is concomitant with a large drop in the piezoelectric response~\cite{Li2016}.  The series of temperature scales characterizing the changes in bulk properties of the relaxors has been interpreted in terms of random field models~\cite{Relaxing,StockSoftModePZN, Westphal, Fisch,Glinchuk}, in analogy to model magnets.  In the case of relaxors, random dipolar fields are introduced through the disorder on the B-site in the ABO$_{3}$ structure where in model magnets random fields are introduced through application of a magnetic field (known as the `Aharony trick'~\cite{Fishman79:12}).

Similar to model random field magnets, relaxors have also been shown to display a ``skin effect" where near-surface regions (where the unit cell shape differs from the bulk) have been directly observed using diffraction techniques in lead based relaxors including PZN, PMN and their PT doped variants~\cite{SkinPMNPT1, SkinPMNPT2, SkinPZN, SkinPZNPT, SkinPMN1, SkinPMN2,Skinreview,Phelan15:88}.  This has also been demonstrated in lead-free relaxors~\cite{Kong19:29} illustrating the commonality of this macroscopic skin region in disordered relaxors.  Depth dependent chemical analysis in large single crystals have suggested the relative concentration on the B-site to be a possible origin~\cite{BrownPMN} which would in turn lead to a change in the random fields.  However, recent depth dependent X-ray diffraction~\cite{Kumar20:8} has pointed to relative oxygen concentration as the origin of this skin-region and indeed have been able to tune dielectric properties through oxygen doping\cite{Kong19:29}. Chemical homogeneity was actually implicated in the original paper by Smolenski~\cite{Smolenski60} as the origin of the diffuse phase transitions in relaxors.  We note that the observation of a skin region that is physically distinct from the bulk have been made in free-standing single crystals in a number of compounds and has been referred to as the two-lengthscale problem.~\cite{Cowley96:66}  Examples include rare-earth metals Ho~\cite{Thurston93:70,Gehring93:71,Hirota94:49,Gehring95:51} and Tb, the perovskite insulator SrTiO$_{3}$~\cite{Shirane93:48,Hochli82:48,McMorrow90:76}, and the spin-Peierls compound CuGeO$_{3}$~\cite{Wang01:63}.  Recently, differing magnetic properties at the surface in comparison to the bulk have been reported in van der Waals magnets with Fe$_{1+x}$Te being an example.~\cite{Trainer19:5,Trainer21:103}  However, relaxors are distinguished over these systems with a macroscopic ($\sim$~100~$\mu$m) ``skin" present over a very wide temperature range.  We note that this lengthscale has been directly verified with both X-ray~\cite{Kumar20:8,SkinPMN2,SkinPZN,SkinPZNPT} and neutron diffraction utilizing spatial strain scanning~\cite{SkinPMN2,Skinreview}. 

Whilst there has been some debate in the literature over the relaxor nature of PFN~\cite{raevski2012}, it typically displays broadened dielectric properties similar to those for PMN and PZN~\cite{gridnev2012, brzezinska2018, 10.1063/1.2158131, rao_1995}, undergoes a ferroelectric transition at $\sim$~400 K~\cite{Wilfong16:47, Bonny97:102} and displays spatially short-range polar correlations similar to the polar nanoregions in PMN and PZN~\cite{Burns2}. This suggests that it shares much of the fundamental physics of more standard relaxor materials and also underlies the sensitivity to composition which will be discussed specifically in this paper.

A relatively unique feature of PFN compared to other relaxors is the presence of magnetic Fe$^{3+}$ correlations which are potentially related to the dielectric properties.  This is of interest in the context of the broader field of multiferroics where, conventionally, magnetic and electrical order are both present in the same phase~\cite{Spaldin_Review, Spaldin_Ramesh_Review}. The coupling of these order parameters would allow the design of hybrid magnetoelectric devices that are independently controllable using magnetic and electric fields~\cite{Kimura_TMO}. However, in usual multiferroic materials this coupling is typically weak due to the competing interests of ferromagnetism (requiring partially occupied orbitals) and ferroelectricity (preferring filled orbitals)~\cite{Hill}. 

PFN combines both ferroelectric and magnetic properties and so is classed as a multiferroic.  Evidence for this has been reported~\cite{YangPFN, StockPFN} along with exceptionally large magnetoelectric coupling~\cite{Laguta16:51,Laguta17:95}. PFN contains a mixture of Fe$^{3+}$ (magnetic with $S=5/2$, $L\approx0$) and Nb$^{5+}$ (non-magnetic) on the B-site~\cite{Maryanowska}. Ferroelectric order has been observed below $T_{\textrm{C}} \approx$~400~K, evidenced through neutron spectroscopy by the corresponding low temperature energy hardening of a zone center transverse optic soft mode~\cite{StockPFN}.  Unlike classic ferroelectrics~\cite{ShiraneSoftModePT, KempaSoftModePT} and non-magnetic relaxors~\cite{WakSoftModePMN, StockSoftModePZN} where the soft phonon ($\hbar \Omega_{0}$) driving polar correlations scales as $(\hbar \Omega_{0})^{2}\propto (T-T_{\textrm{C}})$, the energy of the soft mode in PFN deviates from this mean-field result at low temperatures.  Based on sum rules of magnetic neutron scattering, this was suggested to result from the development of spatially short-range magnetic correlations and indicative of an underlying coupling between magnetic and polar orders~\cite{StockPFN}.

PFN has been reported to undergo two magnetic phase transitions/anomalies with the first at $T_{\textrm{N}}\approx$~150~K~\cite{Bokov2,Howes} which corresponds to the formation of spatially long-range antiferromagnetic ordering of Fe$^{3+}$ spins and the second below $T_{\textrm{g}} \approx$~15-20~K which corresponds to a glass transition~\cite{Blinc, Bhat} where zero and field cooled susceptibilities diverge.  Single crystal neutron diffraction measurements indicate that in PFN samples that display antiferromagnetic order, the Fe$^{3+}$ spins orders as a G-type antiferromagnet characterized by a resolution-limited (in momentum) Bragg peak measured at $\vec{Q} = (1/2,1/2,1/2)$~\cite{Howes}, indicative of spatially long-range magnetic correlations. But, single crystal~\cite{RotaruPFN}~(Fig.~4) and magnetic powder work~\cite{Pietrzak81:65} have shown that such spatially long-ranged magnetic order is not complete and coexists with short-range correlations in materials where single crystal neutron diffraction data has been reported. Of particular relevance is powder diffraction data showing only a fraction ($\sim 3$~$\mu_{B}$) of the full magnetic moment ($gS = 5$~$\mu_{B}$) accounted for in the magnetic Bragg peaks.~\cite{Ivanov00:12} This is also consistent with NMR work on Fe$^{3+}$ based relaxors.~\cite{Zagorodniy18:2}  At low temperatures below $T_{\textrm{g}}$, a cluster spin-glass phase~\cite{Kumar} coexists with the previous static long/short-range magnetic correlations~\cite{KleemannPFN,RotaruPFN}.

The details of the magnetic properties in Fe$^{3+}$ relaxors has been shown to be sensitive to the relative Fe$^{3+}$ concentration.   While the cluster spin-glass transition have been found in all samples (along with ferroelectric transitions), where reported, the N\'eel transition~\cite{Pietrzak81:65} at T$_{N}$~$\sim$~150~K is not universal across samples~\cite{Raevski}.  In particular, ~\cite{Zagorodniy18:2} report a study of the magnetic properties of PFN and also Pb(Fe$_{1/2}$Sb$_{1/2}$)O$_{3}$ applying NMR and find only partial magnetic order.  This is further illustrated in~\cite{Bochenek18:11} who report the magnetic properties in PFN synthesized with differing relative Fe$^{3+}$/Nb$^{5+}$ concentrations with the antiferromagnetic peak in the susceptibility \textit{not} being observable for reduced iron concentrations.  The spatially long-ranged N\'eel phase is not universal to all PFN samples and given the issue of random fields in controlling the relaxor properties (which share the compositional disorder), this needs to be characterized in detail in large single crystals.

This paper examines the magnetism in a large (1~cm$^3$) single crystal sample of PFN grown with the modified-Bridgman method~\cite{Luo00:39}.  We characterize the low frequency and nearly static magnetic properties using muon spectroscopy.  By tuning the momentum of the incident muons, we control the implantation depth of the muons, thereby characterizing the slow magnetic fluctuations as a function of depth.  We find a change in the magnetic fluctuations over the same lengthscale reported for ferroelectric skin effects in relaxors.  We therefore report a magnetic skin effect in PFN.

This paper is split into three further sections. \Sref{sec:methods} describes the two experimental muon techniques that were used in this paper. This is followed by \sref{sec:results} which outlines the results - \sref{sec:MuSR} for the positive muon ($\mu$SR) experiment and \sref{sec:NegMu} for the negative muon compositional analysis experiment. Lastly, we present a discussion of the results in \sref{sec:disc}.


\section{Experimental methods}\label{sec:methods}

In this section we discuss the materials characterization of our PFN sample followed by a description of the muon experiments.

\subsection{Materials}\label{sec:materials}

As outlined above, the magnetism in disordered Pb(Fe$_{1/2}$Nb$_{1/2}$)O$_{3}$ has been reported to have a variety of responses. Therefore, we outline the synthesis and characterization in our 1~cm$^{3}$ single crystal used in both the muon measurements discussed here and in~\cite{neutronPaper,StockPFN}. The single crystal was grown using the modified-Bridgman technique (outlined in~\cite{Luo00:39}) following the procedure in~\cite{Kozlenko14:89}. 

A small piece taken from our large single crystal was previously used to perform Raman spectroscopy~\cite{Wilfong16:47} as a function of pressure where it was found to undergo two structural transitions consistent with X-ray diffraction reporting the same ferroelectric transitions~\cite{Bonny97:102}.  This is also consistent with a softening of the optical transverse phonon mode at the nuclear zone center reported using neutron spectroscopy~\cite{StockPFN} which is indicative of ferroelectric order in perovskites.  As a further check of the sample, we performed single crystal neutron diffraction where the Fe and Nb concentrations refined to 0.5:0.5, within error.  We note that the structural and ferroelectric transitions in our sample are consistent with those reported previously.  

The temperature dependent magnetic susceptibility in our sample has been reported previously in~\cite{StockPFN} where a low temperature ($\sim$~15-20~K) anomaly was found with differences between field and zero-field cooled behavior observed.  This is consistent with other reports of the low-temperature susceptibility such as in~\cite{Falqui05:109,KleemannPFN,Laguta14:16,Laguta17:95}.  At higher temperatures, a deviation from the Curie-Weiss law in the magnetic susceptibility is observed below~$\sim$~150~K~\cite{StockPFN}.  This contrasts with some other reports of a peak in the magnetic susceptibility at the same temperature (example in~\cite{KleemannPFN,Laguta14:16}) taken as evidence for long-range antiferromagnetic order.  However, neutron diffraction~\cite{Chillal13:87} found that the low temperature magnetic neutron response had two components with a momentum resolution-limited Bragg peak and a momentum broadened component.  The former Bragg peak corresponds to magnetic correlations that are spatially long-ranged on the resolution of neutron diffraction, while the later momentum broadened peak corresponds to only spatially short-range order.  This is different than a simple bulk phase antiferromagnetic transition and implies two components at low temperatures.  The peak in susceptibility is also not universally reported in the literature with~\cite{Bochenek18:11} showing a wide variety of responses depending on relative Fe$^{3+}$/Nb$^{5+}$ concentrations.  A comparison amongst a number of Fe$^{3+}$ lead-based relaxors using NMR was reported in~\cite{Zagorodniy18:2} and found that this peak was strongly related to the disorder on the B-site as well as confirmation of multiple Fe$^{3+}$ sites.  However, consistent with the literature and the discussion above, it was found that the ferroelectric transition and the low temperature magnetic susceptibility anomaly (tied with a cluster spin-glass transition) to be robust.  

To summarize, the single crystal discussed here displays the universal low-temperature glass anomaly in the susceptibility and the ferroelectric transitions at high temperature consistent with previous samples reported in the literature.  An observable peak in the susceptibility at $\sim$~150~K is absent in our sample while only a deviation from the Curie-Weiss law is found.  However, this susceptibility peak is not universally reported in the literature and is found to be highly dependent on disorder on the Fe$^{3+}$ site and also found to coexist with spatially short-range magnetic order where single crystal neutron diffraction work has been reported.

\subsection{Muon spectroscopy}

In this paper both positive ($\mu^{+}$) and negative ($\mu^{-}$) muons are used to characterize a $1 \times 1 \times 1$~cm$^3$ cube single crystal of PFN. All experiments were carried out at the RIKEN-RAL facility (ISIS, STFC Rutherford Appleton Laboratory, UK) on CHRONUS ($\mu^{+}$) and MuX ($\mu^{-}$).

Firstly, the positive muon spin relaxation ($\mu$SR) technique was used to probe the internal magnetic dynamics of PFN~\cite{Dalmas_MuSR_Ref, Blundell_MuSR_Ref}. This technique uses a polarized beam of $\mu^{+}$ which will show spin precession under a uniform magnetic field. Upon generation, the muon beam at ISIS is polarized (due to the parity-violating electro-weak pion decay process) and, when the $\mu^{+}$ subsequently decay into positrons ($e^{+}$), they are preferentially emitted along the direction of the spin (again due to electro-weak interaction). As the spin of a $\mu^{+}$ will precess under the influence of a magnetic field, by detecting the direction of the $e^{+}$, this precession can be measured. As the precession frequency is directly proportional to the field strength, this measurement can be used to determine the magnetic field at the site of muon implantation (the incident polarization direction of the muon is known due to the polarized beam). 

In a $\mu$SR experiment, this directional anisotropy of the muon decay is characterized by a quantity known as the asymmetry, $A$. This is defined as the normalized difference between the number of $e^{+}$ detected in the forward ($F$) and backwards ($B$) directions

\begin{equation}
	\fl A = \frac{F - \alpha B}{F + \alpha B},
\end{equation}

\noindent where the parameter $\alpha$ corrects for the difference in efficiency between the two detectors. This normalized difference removes the characteristic decay curve of the muons which would otherwise be superimposed on the data. The sample was wrapped in Ag foil for this experiment and placed in a Janis dynamic He flow cryostat with aluminum windows for beam access.   We note the use of Ag foil owing to the fact that has been measured not to depolarize the muon beam~\cite{Bueno11:83} and slits were used to mask the sample holder so that the muon beam only is incident on the PFN sample and surrounding Ag foil.   

Secondly, the $\mu^{-}$ technique provides a method to analyze the composition of a material in a non-destructive way. In this technique~\cite{Hillier,Lin23:19}, $\mu^{-}$s are implanted into the sample where they interact locally with an ionic site.  After muon capture onto the valence band, the $\mu^{-}$ cascades down the (modified) atomic orbitals emitting a characteristic X-ray spectrum which allows the ionic site to be identified. Due to the larger mass of the $\mu^{-}$ (105.7~MeV/$c^2$), these muonic X-rays show less re-absorption by the material than if this experiment were carried out with electrons (0.511~MeV/$c^2$).

The $\mu^{-}$ experimental setup consists of four detectors (two upstream and two downstream) placed on a flat surface with the sample held in the center. The sample was held in an Al foil packet and suspended into the beam.

Being a spallation source, the muons at ISIS are produced by the interaction of an accelerated proton beam with a graphite target~\cite{RAL}. This reaction consumes approximately 5\% the incident proton beam and produces pions which in turn decay into muons. Each spill from the accelerator produces two pulses of pions (separated by 320~ns) with each pulse having an FWHM of 55~ns~\cite{muonsISIS}. Depending on the energy of the pion when it decays, it can produce either `surface' or `decay' muons. The former are from pions that form at rest (and so on the \emph{surface} of the target) and are only capable of producing $\mu^{+}$ (due to the re-capture of $\mu^{-}$) whilst the latter form from pions that decay in the beamline and can produce both $\mu^{+}$ and $\mu^{-}$~\cite{muonsISIS}.

Whereas the other beamlines at ISIS operate at a fixed momentum (28~MeV/c), the RIKEN-RAL beamline contains an array of magnets which allow the momentum of the muon beam to be tuned to the experimental need. These magnets also enable pions to be extracted from the muon target and thus allow the formation of decay muons in the beamline. Hence, RIKEN-RAL is capable of production of both $\mu^{+}$ and $\mu^{-}$ at a range of momenta~\cite{muonsISIS}. Given that the key advantage of muon techniques is that they are local probes (the $\mu^{-}$ due to its sensitivity to the ionic structure at the implantation site and the $\mu^{+}$ to the local magnetic field), varying the momentum of the muon beam allows the implantation depth to be controlled and thus provides a method for depth dependent magnetic analysis. Due to the depth profile required for this experiment, other sources of low energy muons such LEM at PSI (Villigen, Switzerland) are unsuitable; RIKEN-RAL is the only beamline that fulfills our requirements.

In this way both the spatially-local magnetic field and composition of PFN were profiled against depth. Whilst decay muons were necessarily used for the $\mu^{-}$ elemental analysis, surface muons were used for the $\mu$SR experiment as, in the momentum range required to achieve desired penetration into the sample, they have a much higher production rate. However, this also limited the upper momentum to 27~MeV/c as, above this, the intensity quickly drops by several orders of magnitude~\cite{RIKEN, muonsISIS}.


\section{Results}\label{sec:results}

This section presents results from two series of muon experiments that studied the magnetic properties (\sref{sec:MuSR}) and composition (\sref{sec:NegMu}). Both sets of experiments investigated the properties as a function of muon momentum tuned with the goal of investigating depth dependent properties of the sample.  For both experiments the $\mu^{\pm}$ beam was incident onto a [100] crystallographic face.  We first present the results of both $\mu^{+}$ and $\mu^{-}$ experiments followed by the discussion.

\subsection{Muon spin relaxation with $\mu^{+}$}\label{sec:MuSR}

In order to study the magnetic properties, the $\mu$SR technique was used with $\mu^{+}$. In this section we first describe zero field $\mu$SR measurements as a function of depth, tuned with the incident muon momentum. We also confirm the lack of strong magnetic order using longitudinal fields. Finally we discuss on the temperature dependence of the stretched exponential parameters under warming.

\subsubsection{Depth study}\label{sec:ZF}

\begin{figure}[t]
	\centering
	\includegraphics[width=.65\linewidth]{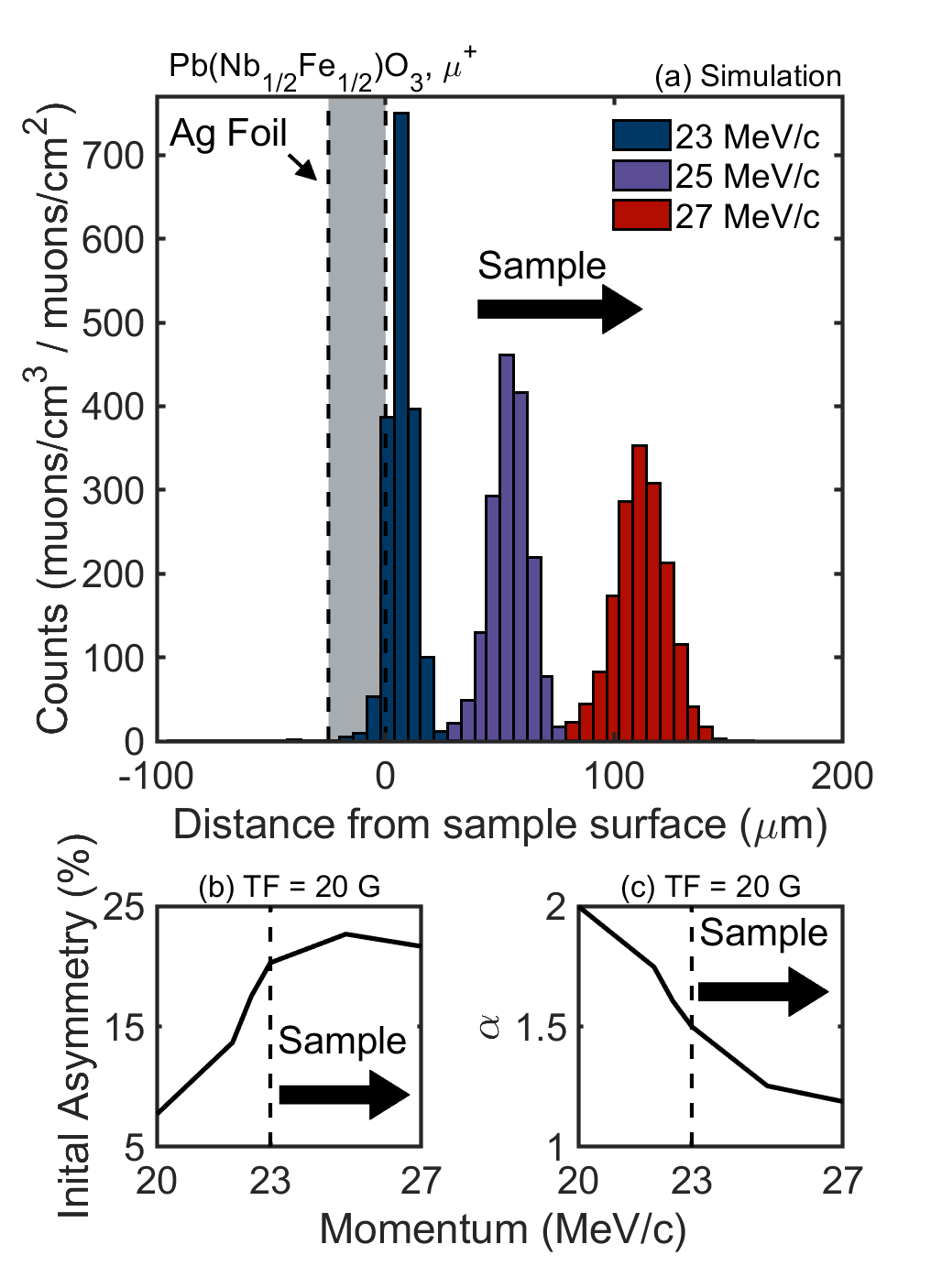}
	\caption{(a) Shows the results of a simulation in \textsc{SRIM/TRIM}~\cite{srimtrim} of $\mu^+$ implantation with momenta of 23~MeV/c, 25~MeV/c and 27~MeV/c. The peak of the distributions give the implantation depths as $\approx$~5~$\mu$m, $\approx$~50~$\mu$m and $\approx$~110~$\mu$m respectively. The y-axis units correspond to implantation density (muons/cm$^3$) per unit flux through the implantation surface (muons/cm$^2$). Also shown are the measured initial asymmetry (b) and $\alpha$ (c) values plotted against depth for a 20~G magnetic field orientated in the direction transverse to the muon spin polarization measured on CHRONUS. The drop in asymmetry below 23~MeV/c indicates that the surface is near here which is in excellent agreement with the simulation.}
	\label{fig:charac}
\end{figure}

Firstly, experimental runs at different momenta (implantation depth) were undertaken in order to allow the local magnetic properties to be profiled against depth in our PFN single crystal. Muon implantation at a given momentum was simulated in \textsc{SRIM/TRIM}~\cite{srimtrim} and the results of these calculations are shown in \fref{fig:charac}~(a). For the purpose of this simulation, the
muon was treated as a modified H$^{+}$ ion (being approximately nine times lighter than
the proton). The experimental setup was reconstructed in the simulation which allowed the stopping profile of the $\mu^{+}$ to be determined for a given momenta. It was found that, for the $\mu^{+}$ experiment, 23~MeV/c corresponded best to an implantation depth of $\approx$~5~$\mu$m, indicated by the maximum of the distribution (blue in \fref{fig:charac}~(a)). The maximum momentum of 27~MeV/c is also shown (red) and corresponds to an implantation depth of $\approx$~110~$\mu$m. As mentioned above in \sref{sec:methods}, any momentum above 27~MeV/c would require a change from surface to decay muons to avoid a large drop the $\mu^{+}$ production rate and this was not feasible due to beam time constraints. This penetration range overlaps with the lengthscales observed for the skin effect in non-magnetic relaxors~\cite{SkinPZN,SkinPMN2,SkinPZN3,Skinreview}.

\begin{figure}[t]
	\centering
	\includegraphics[width=.65\linewidth]{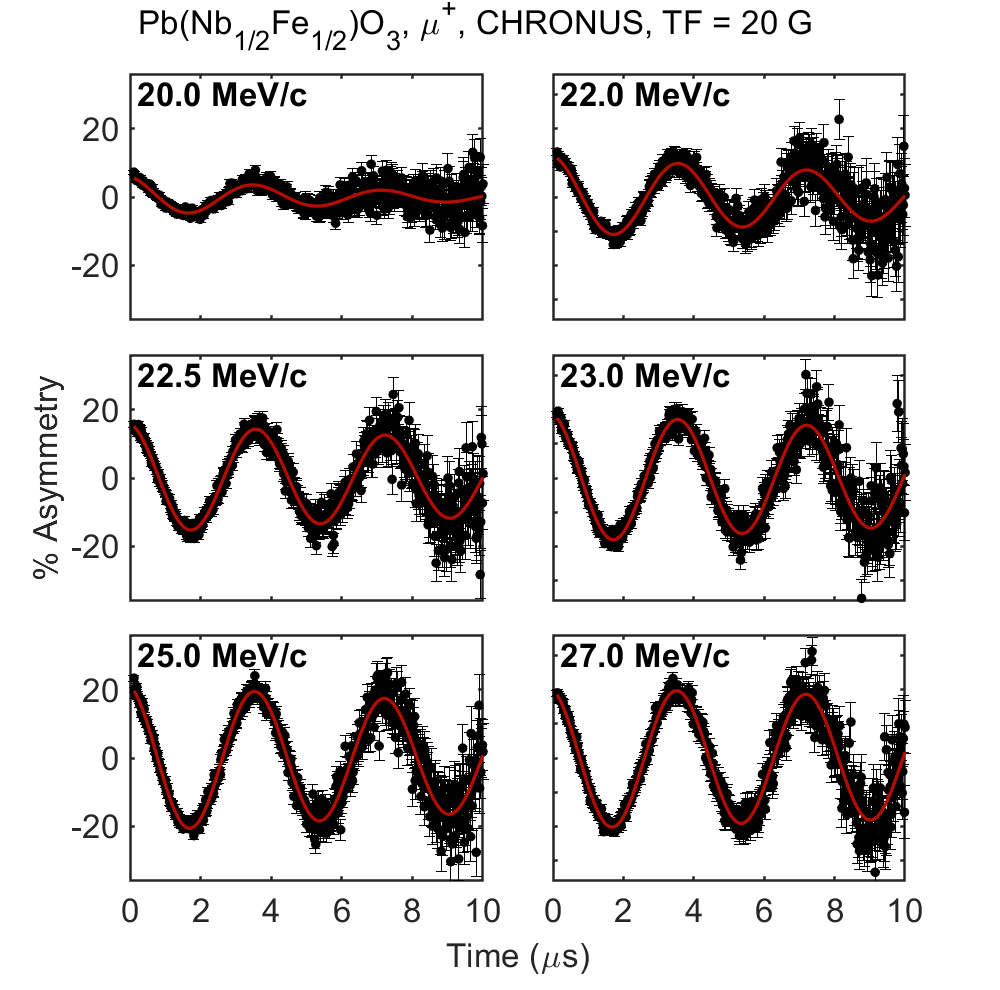}
\caption{Muon asymmetry against time showing the $\mu$SR response under transverse field (TF) conditions against momentum (depth) for a field = 20~G at room temperature. Oscillating fit is shown in red. The drop-off in amplitude below 23~MeV/c indicates that this is the surface region in agreement with the simulations shown in \fref{fig:charac}~(a).}
\label{fig:TF}
\end{figure}

Characterization of the experimental setup was carried out under an applied field of 20~G orientated transverse (TF) to the direction of spin polarization (as standard). This allowed the value of the forward-backward efficiency correction $\alpha$ and the initial values of the asymmetry to be extracted for each momentum. To achieve this the forward and backward detector groups were fitted separately in \textsc{WiMDA}~\cite{WiMDA_ref} and these were combined to produce a fitted asymmetry profile. The data have been corrected for the double pulse structure of the ISIS beam by implementation of a time offset of 244~bins (1~bin~=~16~ns) such that the fitted range only corresponds to the sample response.

A (relatively strong) transverse field ensures that there will be measurable muon precession signal and the amplitude of this gives an estimate of the initial asymmetry. The results of this are shown in \fref{fig:TF}. The full asymmetry on CHRONUS (at time of measurement) is $\approx$~23\%. The measured values of initial asymmetry and $\alpha$ are plotted in \fref{fig:charac}~(b) and~(c) respectively against momentum. These are taken from the fits shown in \fref{fig:TF} in red. The crystal surface corresponds to the momentum where the asymmetry drops below 23\% because it indicates a reduced sample implantation. This agrees with the simulations shown in \fref{fig:charac}~(a). A further \textsc{SRIM/TRIM} simulation confirmed that for momenta below 23~MeV/c the majority of muons do not stop in the sample or Ag foil rather in the cryostat. We speculate that the depolarization observed could be the result of muonium formation in the mylar layer of the cryostat~\cite{Yaouanc_muon_2011}. Based on these preliminary measurements, the momenta for the main experiment were chosen as 23~MeV/c, 25~MeV/c and 27~MeV/c.

\begin{figure}[t]
	\centering
	\includegraphics[width=.65\linewidth]{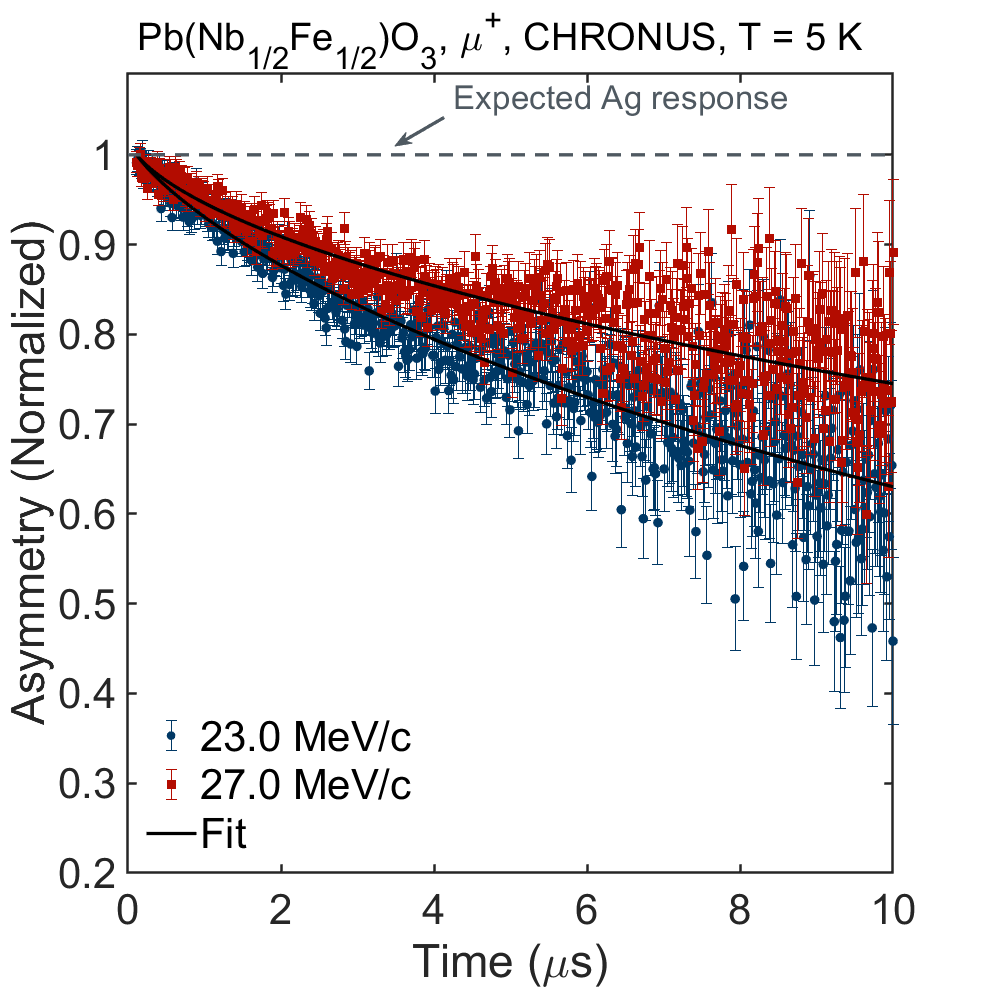}
\caption{Muon asymmetry against time showing the $\mu$SR response under zero field (ZF) conditions. Plotted are the shallowest (23~MeV/c, blue) and deepest (27~MeV/c, red) depths to highlight the difference in the muon spin relaxation as a function of depth. Also plotted are the stretched exponential fits (solid lines) that are discussed further in the main text.  The expected response for silver is also plotted taken from~\cite{Bueno11:83}.}
\label{fig:ZF_curves}
\end{figure}

The main experiment was carried out under zero field (ZF) conditions at a temperature of 5~K. The data for two extremes of muon momentum (corresponding to `surface' and `bulk' measurements) are compared in \fref{fig:ZF_curves} where a slowing down of the relaxation with depth can be seen. The data were fitted using the stretched exponential function

\begin{equation} \label{eq:asym_Eqn}
	\fl A(t) = A_0 e^{-(\lambda t)^{\beta}} + B_0,
\end{equation}

\noindent where $A_0$ is the amplitude of the (normalized) relaxing asymmetry, $\lambda$ is the effective depolarization rate, $\beta$ is a power exponent and $B_0$ is a background term. The use of this function and its relation to the local magnetic field distribution will be discussed later, but it allows effective parameterization of the asymmetry decay curve. 

The data were normalized using the initial asymmetry values taken from the preliminary TF runs i.e. the amplitude of the oscillations in \fref{fig:charac}~(b). The period of oscillation is determined by the applied field. This normalization allows meaningful comparisons between data at different momenta as the change in implantation depth will affect the value of $\alpha$ which has the effect of shifting the relaxation curve. However, it is important to note that the relaxation parameters $\lambda$ and $\beta$ are unaffected by this correction. 

Although, we note that the measured relaxation response differs from what is expected in Ag~\cite{Bueno11:83} used to mount the sample (which shows no relaxation), the effect of $\mu^{+}$ stopping short must be accounted for, especially near the surface. The proportion of $\mu^{+}$ which stop in Ag was calculated from the stimulated stopping profiles shown in \fref{fig:charac} and found to be 11.5\% at 23~MeV/c dropping to 0.33\% at 25~MeV/c and 0.04\% at 27~MeV/c. This was implemented into the fitting procedure through a constraint on the value of $B_0$.  

\begin{figure}[t]
	\centering
	\includegraphics[width=.65\linewidth]{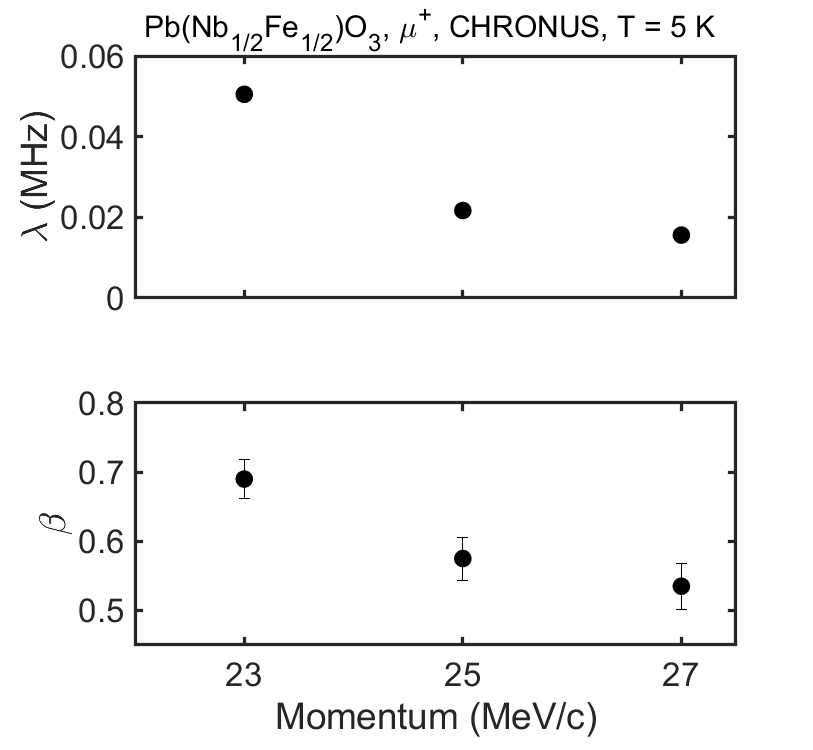}
\caption{Values of relaxation rate, $\lambda$, and exponent, $\beta$, plotted against muon momentum (depth). Both $\lambda$ and $\beta$ decrease with depth. The error bars are smaller than the symbols used to plot the data.}
\label{fig:ZF_depth}
\end{figure}

The fitted values for the relaxation parameters $\lambda$ and $\beta$ are plotted against muon momentum (and hence varying depth) in \fref{fig:ZF_depth}. It can be seen that both $\lambda$ and $\beta$ decrease with depth. This would indicate there is indeed a change in the magnetic properties as the depth is increased. We discuss what these measurements correspond to in the frequency domain in the discussion below.

\subsubsection{Longitudinal field study}

\begin{figure}[t]
	\centering
	\includegraphics[width=.65\linewidth]{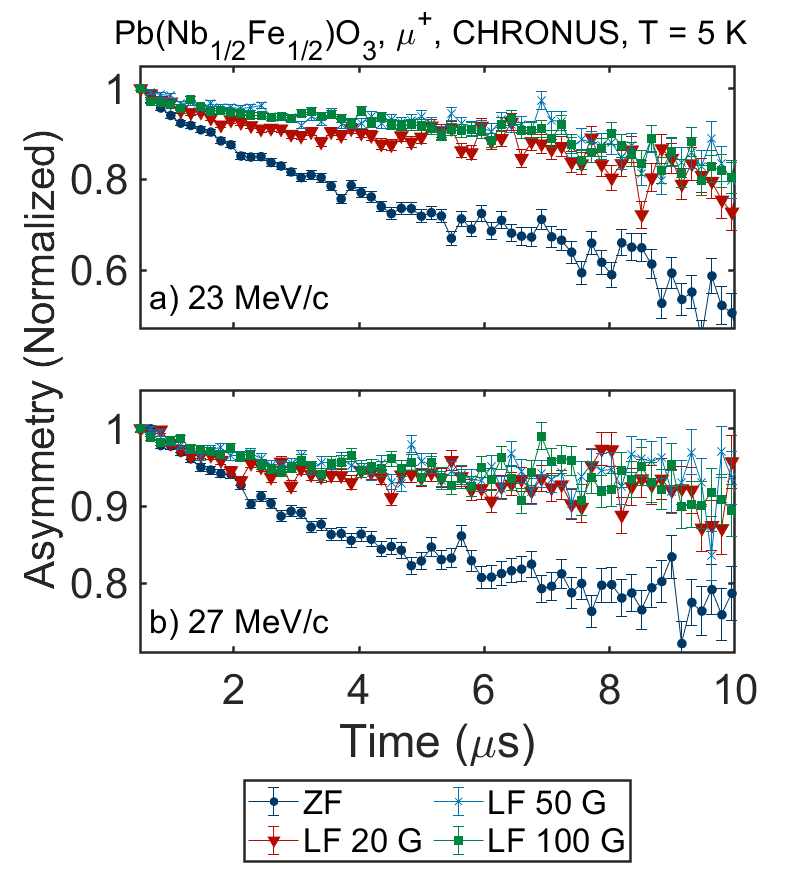}
\caption{Muon asymmetry against time showing the $\mu$SR response under longitudinal field (LF) conditions at a momentum of (a) 23~MeV/c and (b) 27~MeV/c. Magnetic field strengths from 0~G (ZF) to 100~G are shown. A bunching factor (average) of 5 was used in \textsc{WiMDA}~\cite{WiMDA_ref} and the data have been corrected for a field dependent $\alpha$.}
\label{fig:LF}
\end{figure}

With the absence of any observable long-range order in the ZF measurements, a longitudinal field (LF) study was used to characterize the local magnetic field in our PFN sample. Figure~\ref{fig:LF} shows the $\mu$SR relaxation under LF conditions for various $\mu^+$ momenta. In this setup the magnetic field is oriented parallel to the initial $\mu^+$ spin polarization direction and so, under a large enough field, the spin of the $\mu^+$ will align to this rather than the local magnetic structure of the sample. This causes the asymmetry to remain near unity. Due to this external coupling being parallel to the measurement direction, the forwards-backwards correction factor is affected and the data have been corrected for a field-dependent $\alpha$.

The flattening of the relaxation above ZF shown in \fref{fig:LF} indicates that the $\mu^+$ response decouples from the internal magnetic structure of the PFN sample at low fields. Fields up to 500~G were also applied but no changes in the response were seen after 20~G with the asymmetry decoupling towards unity. This was found to be consistent across all depths and such a low decoupling threshold would indicate that the internal magnetic fields are weak. This agrees with the glassy behavior seen in the ZF measurements and the almost immediate decoupling from the internal magnetic field from PFN is consistent with the absence of any spatially long-ranged N\'eel phase. This suggests that the effective moments measured in this sample are small. Furthermore, if the response arose from nuclear moments, these would fluctuate and decouple and, in principle, one would see the relaxation actually being unity with short time fluctuations. The relaxation in \fref{fig:LF} never reaches unity which confirms, along with the depth profiling, that the signal is from magnetic ions. Because the glass transition is slow ($< \mu$s) we would not expect to see any contribution from any time variations during each snapshot which provides our $\mu$SR distribution at low temperature.

\begin{figure}[t]
	\centering
	\includegraphics[width=.65\linewidth]{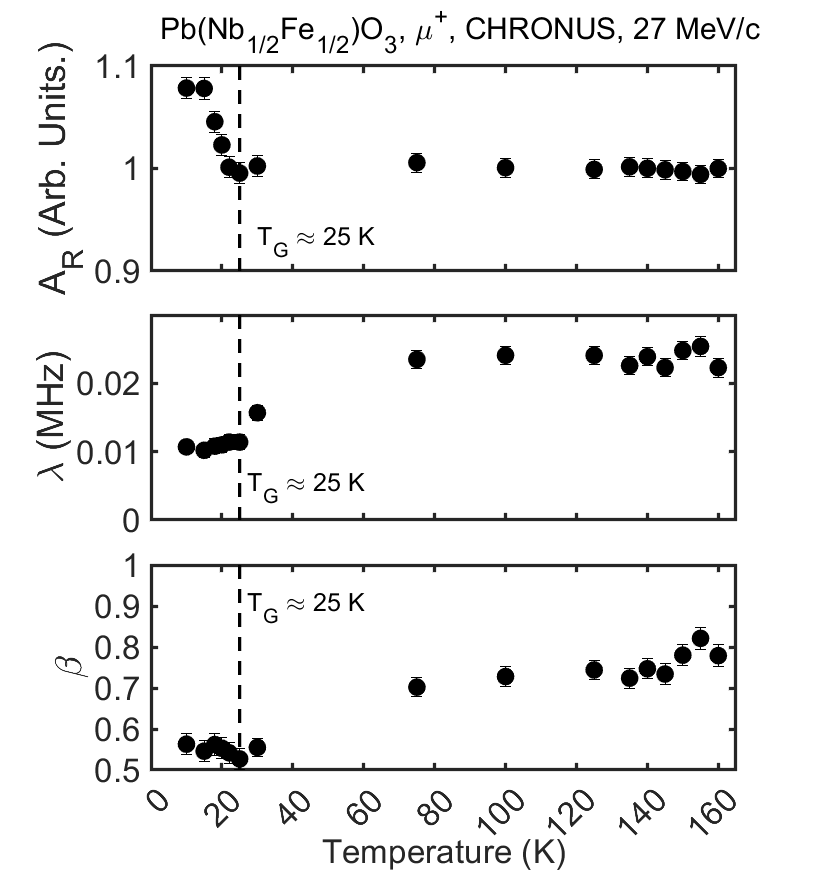}
\caption{Values of (normalized) relaxing asymmetry, $A_R$, relaxation rate, $\lambda$, and exponent, $\beta$ plotted against temperature. The reduction in $A_R$ and increase in $\lambda$ and $\beta$ with increasing temperature are consistent with a transition from a glassy to paramagnetic phase.}
\label{fig:ZF_temp}
\end{figure}

\begin{figure}[t]
	\centering
	\includegraphics[width=.65\linewidth]{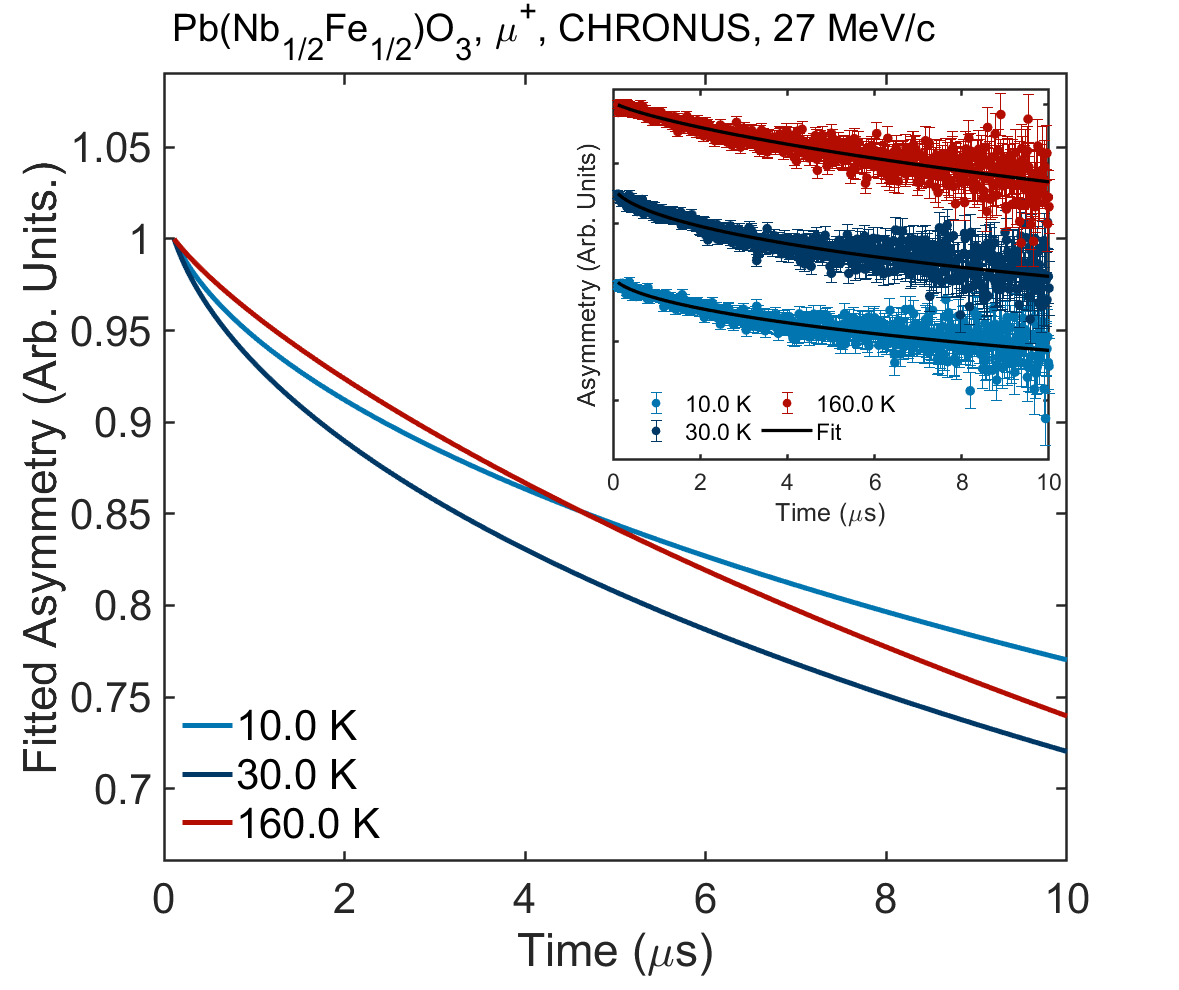}
	\caption{Representative muon relaxation curves at 10, 30, and 160~K.  The fit to a stretched exponential is shown illustrating the differences between the different temperature regions.}
	\label{fig:ZF_Tcurves}
\end{figure}

Lastly, to confirm the presence of the glass transition and the absence of the long-range transition, the longitudinal field was removed and the sample warmed from base to room temperature. The ZF response was measured during the warming and a stretched exponential was used to fit the data. The fitted parameters are shown in \fref{fig:ZF_temp}. Representative relaxation curves over the temperature range are shown in \fref{fig:ZF_Tcurves}. These measurements were done in the bulk (27~MeV/c) as this momentum has the highest $\mu^+$ flux and is representative of the bulk of the sample. Upon removal of the longitudinal field, A$_R$ is increased by about 10\% from the value measured in the depth study presented in \sref{sec:ZF}. This is attributed to the glassy nature of the spin system pining a small fraction of spins along the LF axis with respect to the original distribution.

Step-like changes are observed in all three relaxation parameters near 25~K with a reduction in $A_R$ and increase in both $\lambda$ and $\beta$. This is near the value for the glass transition $T_G$ reported by in~\cite{RotaruPFN} and coincides with temperature anomalies in previously reported magnetic susceptibility measurements~\cite{YangPFN} as well as where zero field and field cooled susceptibility data diverge~\cite{StockPFN} (Figure~1~(d)). This has been interpreted as a transition to glass state, which is consistent with a reduction in the exponent at low temperatures as, in a spin-glass, $\beta \rightarrow {1/3}$ is expected~\cite{Campbell94:72}.  It is also the temperature scale where $\vec{Q}=({1/2},{1/2},{1/2})$ magnetic dynamics are observed to enter the elastic resolution of neutrons~\cite{neutronPaper}.

A glass transition is also consistent with the reduction in $A_R$. As the system is warmed through $T_G$, this extra asymmetry quickly relaxes and $A_R$ then remains at unity (within error) to high temperature. We note that unity here represents agreement with the low temperature depth study as shown in \fref{fig:ZF_curves}. This suggests that the ZF $A_R$ is (within error) independent of temperature.

Furthermore, the N\'eel transition at $\sim$~150~K is not observable in the bulk of our sample from the measurements discussed here.  Such a transition should produce a reduction $A_R \rightarrow 1/3$ at low temperature (as seen in~\cite{RotaruPFN}) - this is not seen in \fref{fig:ZF_temp} with $A_R \approx 1$ further pointing to a lack of a spatially long-range ordered component in our sample in agreement with neutron measurements. 

Finally, the critical-like divergence in $\lambda$ and low temperature recovery of $\beta$ reported in~\cite{RotaruPFN} are absent here. We attribute this to the presence of the remnant field at low temperature which suppresses the relaxation curve as seen in \fref{fig:LF}. The impact of this suppressed relaxation can be seen in the 10~K curve in \fref{fig:ZF_Tcurves}. To truly characterize the static component that would indicate frozen spins, the baseline of the stretched exponential would need characterizing which would require measurements on much longer timescales such as using neutron spin echo.  We discuss the physical connection of these parameters to frequency below after we discuss the stretched exponential in more detail.

\subsection{Compositional analysis with $\mu^{-}$}\label{sec:NegMu}

The momentum dependent study outlined in the previous section indicates a variation as a function of depth sampled through increasing momentum of the incident muons.  Given the discussion in the introduction and the apparent sensitivity of properties in Fe-based relaxors to relative Fe concentration~\cite{Bhat2005, Zagorodniy18:2} and also the existence of a `skin effect' in non magnetic relaxors~\cite{BrownPMN}, we applied negative muon ($\mu^{-}$) spectroscopy to characterize the chemical composition (and particular the relative Fe/Nb ratio) over the same momentum transfer range at room temperature.

\begin{figure}[t]
	\centering
	\includegraphics[width=.65\linewidth]{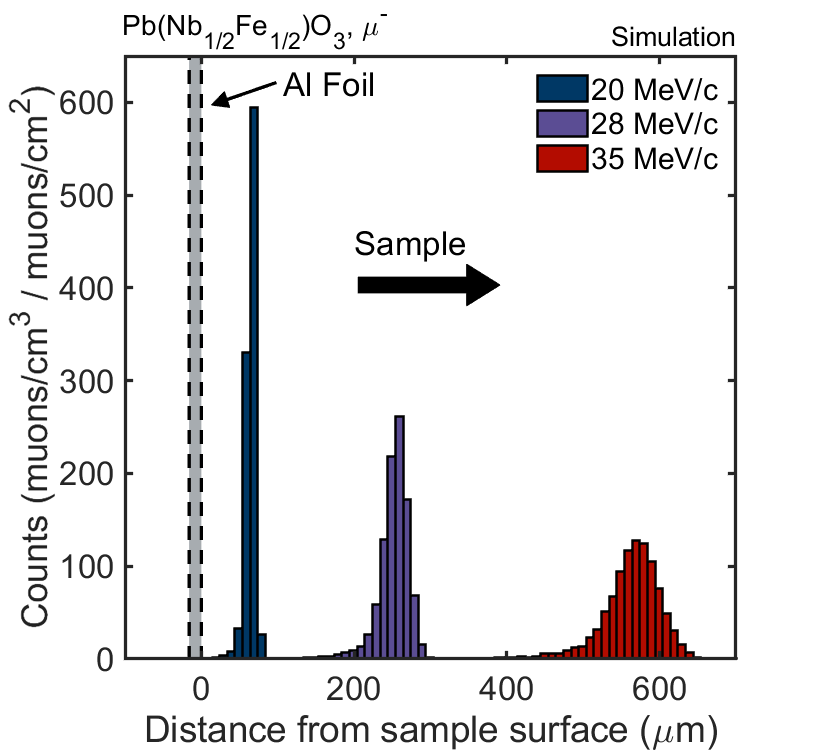}
	\caption{The results of a simulation in SRIM/TRIM~\cite{srimtrim} of $\mu^-$ implantation with momenta of 20~MeV/c, 28~MeV/c and 35~MeV/c. The peak of the distributions give the implantation depths as 65~$\mu$m, 260~$\mu$m and 570~$\mu$m respectively. The y-axis units correspond to implantation density (muons/cm$^3$) per unit flux through the implantation surface (muons/cm$^2$).}
	\label{fig:Mu-DepthsSim}
\end{figure}

Six experimental runs, each at a different muon momentum (corresponding to different implantation depths), allowed the relative concentration of Fe$^{3+}$ and Nb$^{5+}$ to be estimated with respect to depth. Stopping of the muons was again simulated in \textsc{SRIM/TRIM}~\cite{srimtrim}. In this $\mu^{-}$ experiment, the range of muon momenta was 20~MeV/c~-~35~MeV/c and the simulated stopping profiles for three of these momenta are shown in \fref{fig:Mu-DepthsSim}. This shows that 20~MeV/c corresponds to `surface' implantation at $\approx$~65~$\mu$m and 35~MeV/c to `bulk' implantation at $\approx$~570~$\mu$m, as indicated by the maxima of the distributions in \fref{fig:Mu-DepthsSim}. Also shown is the stopping distribution for 28~MeV/c to illustrate the shape of the intermediate profiles. All profiles indicate that no $\mu^{-}$ are stopping in the Al foil packet used to mount the sample. This was confirmed by the lack of any Al signal observed in the measured spectra.

The main X-ray energy range of interest to characterize the relative Fe:Nb ratio for this experiment is 220~-~280~keV and the emission lines contained within were tracked against implantation momentum. These peaks result from muonic X-ray emission by Nb, Pb and Fe and the energies and corresponding atomic transitions are detailed in \tref{tab:Peaks}~\cite{Engfer, MuonWebsite}. This was facilitated by the use of a Ge-based ORTEC (Oak Ridge, TN, US) X-ray detector with range 3~keV~-~1~MeV.

\begin{table}
	\caption{\label{tab:Peaks}Theoretical energies of the muonic transition lines of Pb, Fe and Nb that are relevant to this study~\cite{Engfer, MuonWebsite}.}
 \begin{indented}
	\item[]\begin{tabular}{@{}ccc}\br
		Atom & Atomic Transition & Energy (keV) \\\ns \mr
		Pb & O (6$h$ $\rightarrow$ 5$g$)  & 233.7,  235 \\
		Fe & L (3$d$ $\rightarrow$ 2$p$)  & 265.7, 269.4 \\
		Nb & M (4$f$ $\rightarrow$ 3$d$)  & 231.4 \\ \br
	\end{tabular}
 \end{indented}
\end{table}

\begin{figure*}[t!]
	\centering
	{\includegraphics[width=.95\linewidth]{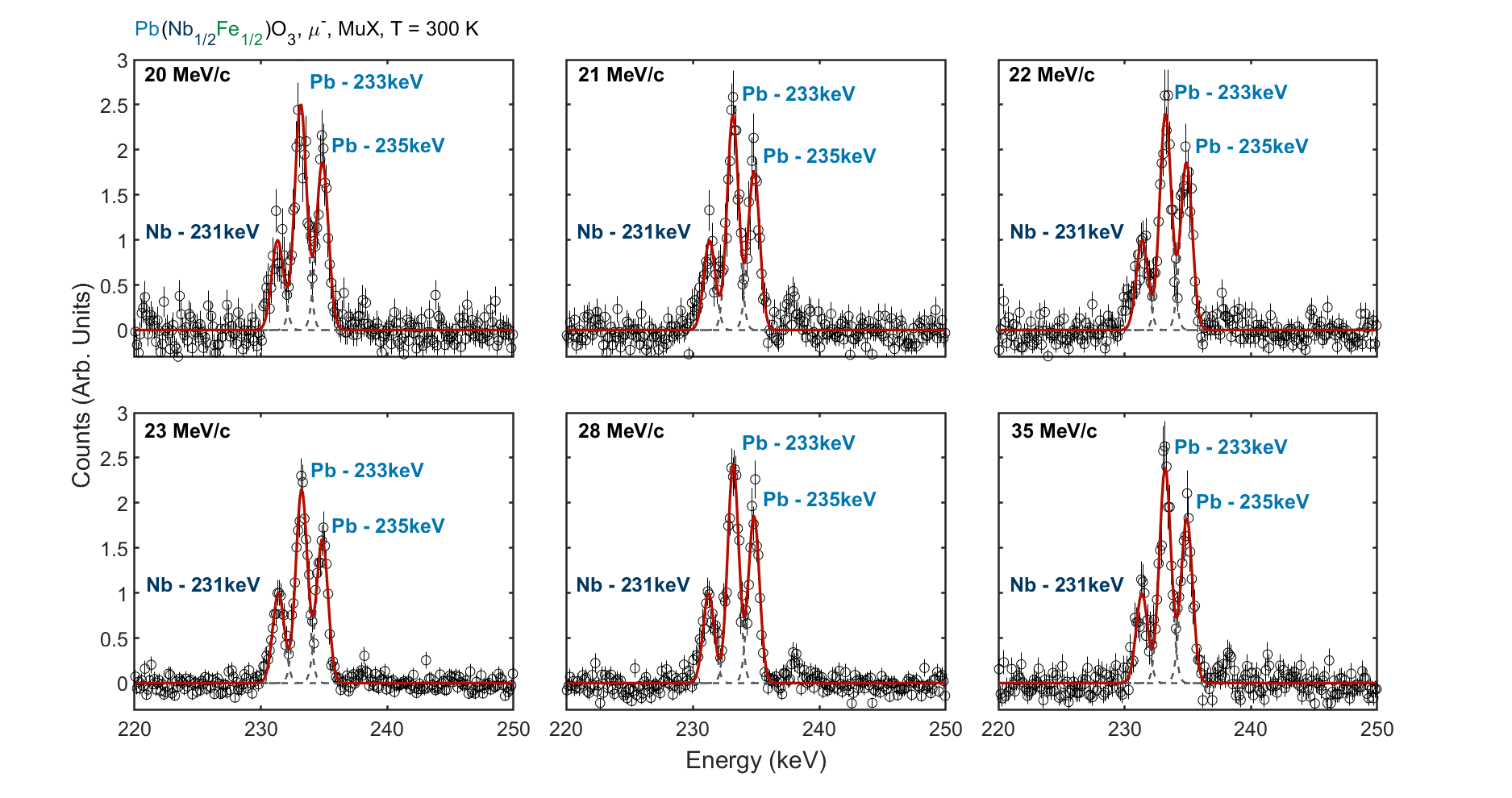}}
	\caption{Muonic X-ray spectra resulting from $\mu^{-}$ implantation into the [001] face of PFN. The panels show different $\mu^{-}$ momenta corresponding to implantation depths between 65~$\mu$m (20~MeV/c) and 570~$\mu$m (35~MeV/c). This energy range shows the Pb and Nb emission lines in \tref{tab:Peaks}. The spectra are normalized with respect to the Nb line at $\approx$~231~keV. The red line shows the fit to the data and the linear background has been subtracted. Errors were estimated using Poisson/counting statistics. The small feature at $\approx$~238~keV is likely another Nb emission line. However, it was not taken into account for this study as it is noticeably weaker than the line at $\approx$~231~keV.}
	\label{fig:spectra1}
\end{figure*}

\begin{figure*}[t!]
	\centering
	{\includegraphics[width=.95\linewidth]{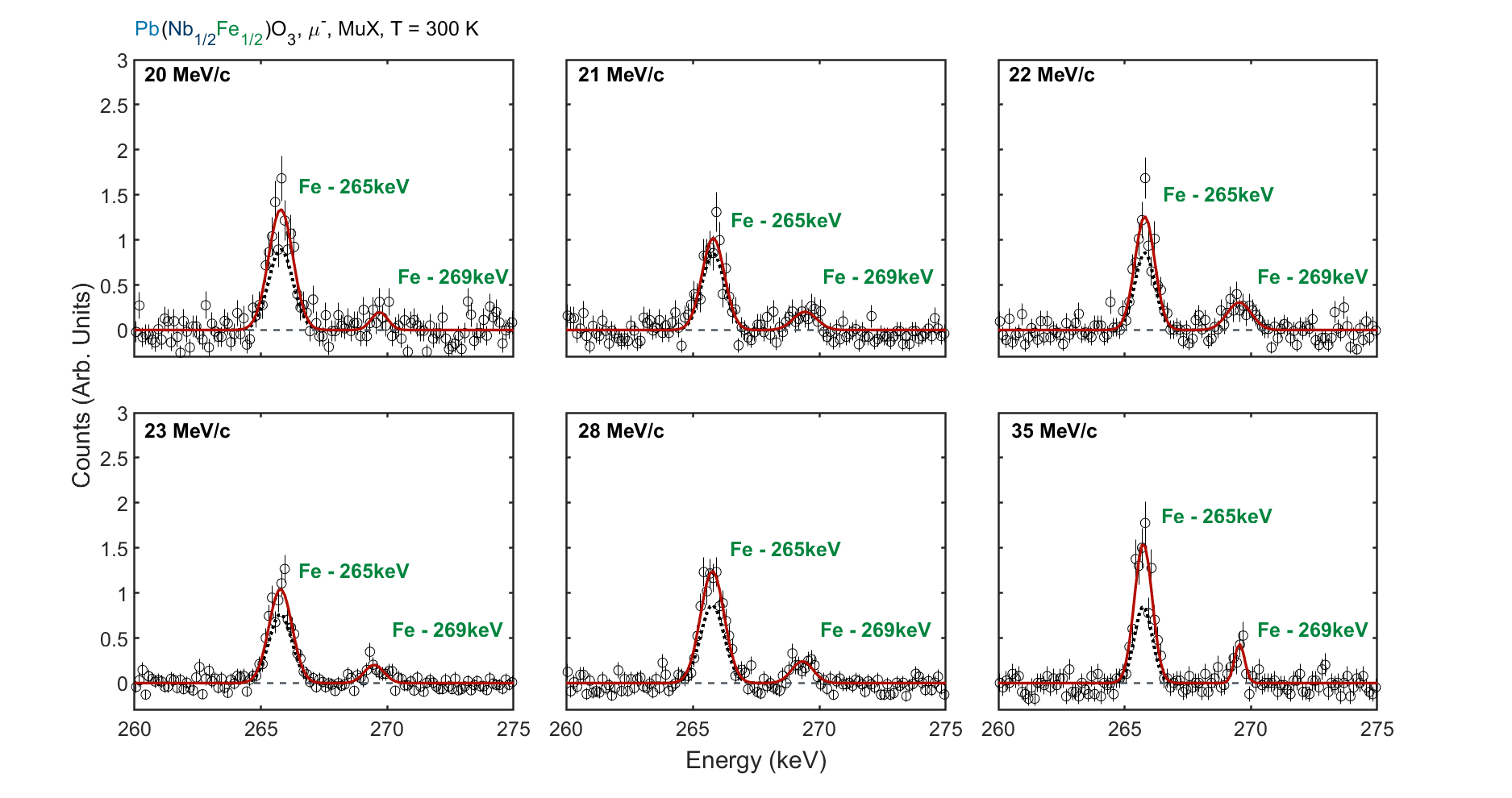}}
	\caption{Muonic X-ray spectra resulting from $\mu^{-}$ implantation into the [001] face of PFN. The panels show different $\mu^{-}$ momenta corresponding to implantation depths between 65~$\mu$m (20~MeV/c) and 570~$\mu$m (35~MeV/c). This energy range shows the Fe emission lines in \tref{tab:Peaks}. The spectra are normalized with respect to the Nb line at $\approx$~231~keV. The red line shows the fit to the data and the linear background has been subtracted. The black dotted line under the 265~keV peak shows the contribution from the $\gamma$ emission from muon capture by the Pb nucleus (discussed in the main text). Errors were estimated using Poisson/counting statistics.}
	\label{fig:spectra2}
\end{figure*}

Figures~\ref{fig:spectra1} and~\ref{fig:spectra2} show the five emission lines resulting from the transitions in \tref{tab:Peaks}. The data were corrected to account for detector efficiency and the errors were estimated using Poisson/counting statistics.  The emission lines were fitted with Gaussian functions and the widths were constrained to be the same for all emission lines consistent with experimental broadening. A linear background term was also included. The spectra are normalized with respect to the Nb line at $\approx$~231~keV to aid in comparison with the Fe lines.

However, analysis of the Fe L muonic X-ray transition lines (at 265.7 and 269.4~keV) is complicated by the presence of a $\gamma$ emission at 265.8~keV which can be attributed to muon capture by the Pb nucleus. In this process the $\mu^-$ is captured by the nucleus following its cascade in a process analogous to electron capture. This results in an excited nucleus of reduced atomic number (Z), which subsequently undergoes $\gamma$ decay. In PFN, a $\mu^-$ may be captured by a Pb nucleus (Z~=~82) which will produce an excited isotope of Tl (Z~=~81). $^{206}$Tl decays from the first excited state to the ground state by $\gamma$ emission with energy =~265.8~keV~\cite{TlGammaTable}.

This $\gamma$ line can also be seen in Pb reference spectrum presented in~\cite{MuonWebsite}, in the spectra reported for PMN in~\cite{BrownPMN} and was also found to be present in unpublished data on the solid solution Pb(In$_{1/2}$Nb$_{1/2}$)O$_3$-Pb(Mg$_{1/3}$Nb$_{2/3}$)O$_3$-PbTiO$_3$ (PIN-PMN-PT) by the authors. All of these samples do not contain Fe and so this peak should represent the pure $\gamma$ contribution. By comparison to these spectra, the intensity of the $^{206}$Tl $\gamma$ emission was estimated to be $\approx$~36\% of the Pb muonic X-ray line at 233.7~keV. For a given $\mu^-$ momentum, these two signals should both be proportional to the abundance of Pb and hence to each other. In this way, an estimate for the contribution of the $\gamma$ at 265.8~keV was included in the fitting of the Fe L doublet.

\begin{figure}[t]
	\centering
	\includegraphics[width=.65\linewidth]{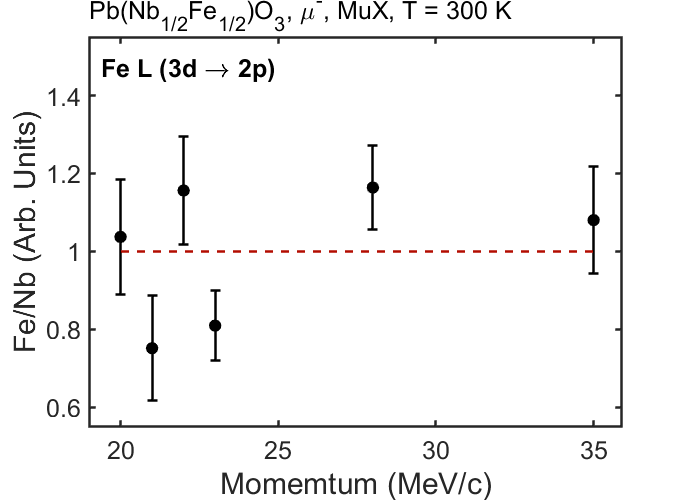}
	\caption{The Fe/Nb fraction against momentum (depth). The figure is scaled such that the average fraction is unity in agreement with the average determined by single crystal diffraction. This fraction was calculated as the ratio of the sum total integrated peak intensity of the Fe L muonic transition lines to the Nb M line.}
	\label{fig:intvsmom}
\end{figure}

\begin{figure}[t]
	\centering
	\includegraphics[width=.65\linewidth]{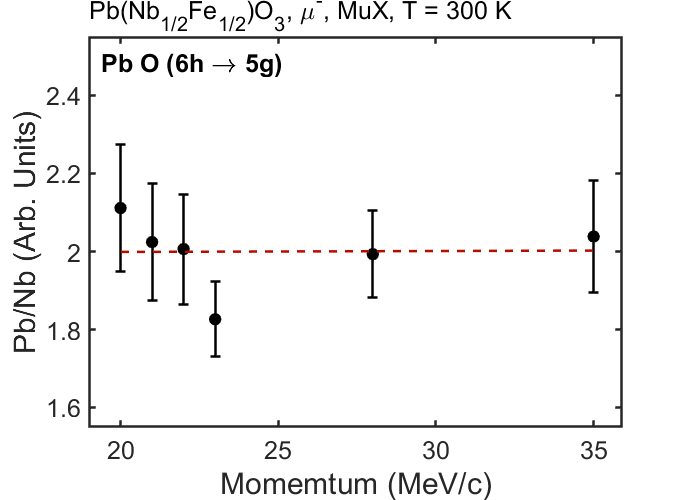}
	\caption{The Pb/Nb fraction against momentum (depth). The figure is scaled such that the average fraction is equal to two in agreement with the average composition determined by single crystal diffraction. This fraction was calculated as the ratio of the sum total integrated peak intensity of the Pb O muonic transition lines to the Nb M line.}
	\label{fig:pb_intvsmom}
\end{figure}

The Fe/Nb fraction was then calculated as the ratio of the total integrated intensities of the two peaks from the Fe L doublet (around $\sim$ 267~keV) and the Nb M line (at 231.4~keV). This is shown in \fref{fig:intvsmom}. Whilst a simple least-squares fit indicates that Fe/Nb fraction decreases by $\approx$~10~-~15\%, the error bars resultant from the fitting procedure are such that care should be taken in such a claim. Hence, on going from the bulk to the surface, we consider the change in Fe/Nb to be within the error of the experiment and so should not be viewed as significant. Hence, within error, we report that the surface composition is approximately consistent with the bulk and therefore that any variation of the Fe/Nb concentration is likely below the order of $\approx$~10~-~15\%. This error could be reduced with a longer counting time but this was not feasible given beamtime constraints. For comparison the change in relative Pb/Nb concentrations is shown in \fref{fig:pb_intvsmom}. The axes of figures~\ref{fig:intvsmom} and~\ref{fig:pb_intvsmom} are scaled such that the average fraction agrees with single crystal diffraction measurement; Fe/Nb~=~1~=~0.5/0.5 and Pb/Nb~=~2~=~1/0.5.  

\section{Discussion}\label{sec:disc}

The $\mu$SR results above indicate a change in the dynamics, parameterized by $\lambda$ in the stretched exponential, as the muons are on average implanted deeper into the sample. To further understand how the dynamics are altered, we discuss below the relation between the stretched exponential in the time domain to the dynamics in frequency which can be directly compared to other probes such as neutron scattering.

\subsection{Stretched exponential in relation to PFN}

Given the complexity of the time structure of the stretched exponential (shown in~\eref{eq:asym_Eqn}) used to parameterize our data, we discuss in detail the structure in the frequency domain.  The use of the stretched exponential can be motivated by a general phenomenological argument: the muon spin experiences exponential relaxation $e^{- s t}$ with the relaxation rate $s$ determined by the strength and fluctuation rate of the local magnetic field experienced by the muon~\cite{Slichter, Uemura}. However, the total $\mu$SR signal will correspond to the spatial average of relaxation rates which is important in the context of PFN that has significant site disorder. The overall relaxation can be modeled stochastically, averaged over a probability distribution $H_{\lambda, \beta}(s)$ based on the implantation site, which mathematically corresponds to the stretched exponential~\cite{stretchedExp}

\begin{equation}
	\fl e^{-(\lambda t)^{\beta}} = \int_{0}^{\infty} ds H_{\lambda, \beta}(s)e^{-st}.
\end{equation}

An integral representation for $H_{\lambda, \beta}(s)$ was first obtained by Pollard~\cite{Pollard} using contour integration (shown in~\eref{eq:PollardForm}). A complementary form~\eref{eq:BSForm} was also derived by Berberan-Santos~\textit{et~al.}~\cite{BerberanSantos1, BerberanSantos2} and this derivation is reproduced in \ref{sec:AppendixA}.  The equivalence of these two forms is explicitly shown in \ref{sec:AppendixA}. However, these integrals are only solvable in terms of standard functions for the cases where $\beta = 1$ or $\beta = 1/2$~\cite{Wagner}. Finally, a series form for $H_{\lambda, \beta}(s)$ is derived in \ref{sec:AppendixB} and shown in~\eref{eq:SeriesForm}. This series was first derived by Humbert and is convergent for all $s$~\cite{Humbert}. It provides an easy computation method~\cite{Steer03:326} and was used to tabulate the distribution by Dishon~\textit{et~al.}~\cite{Dishon}. The normalization of $H_{\lambda, \beta}$ is guaranteed by the assumption that the relaxing asymmetry is likewise normalized - i.e. it starts from unity.

\begin{equation}\label{eq:PollardForm}
	\fl H_{\lambda, \beta}(s) = \frac{1}{\pi \lambda} \int_{0}^{\infty} du \exp\Big({-\frac{su}{\lambda}}\Big) \exp\Big[-u^{\beta}\cos(\beta\pi)\Big] \sin \Big (u^{\beta}\sin(\beta\pi)\Big ).
\end{equation}

\begin{equation}\label{eq:BSForm}
	\fl H_{\lambda, \beta}(s) = \frac{1}{\pi \lambda} \int_{0}^{\infty} dy \exp\Big[-y^{\beta}\cos\big(\frac{\beta\pi}{2}\big)\Big] \cos \Big [y^{\beta}\sin\Big(\frac{\beta\pi}{2}\Big) - \frac{sy}{\lambda} \Big ].
\end{equation}

\begin{equation}\label{eq:SeriesForm}
	\fl H_{\lambda, \beta}(s) = \frac{1}{\pi \lambda}  \sum_{n=1}^{\infty}\frac{\sin(\beta n \pi)}{n!} (-1)^{n+1} \Big(\frac{\lambda}{s} \Big)^{\beta n +1} \Gamma (1 + \beta n). \\
\end{equation}

\begin{figure*}[t]
	\centering
	{\includegraphics[width=1\linewidth]{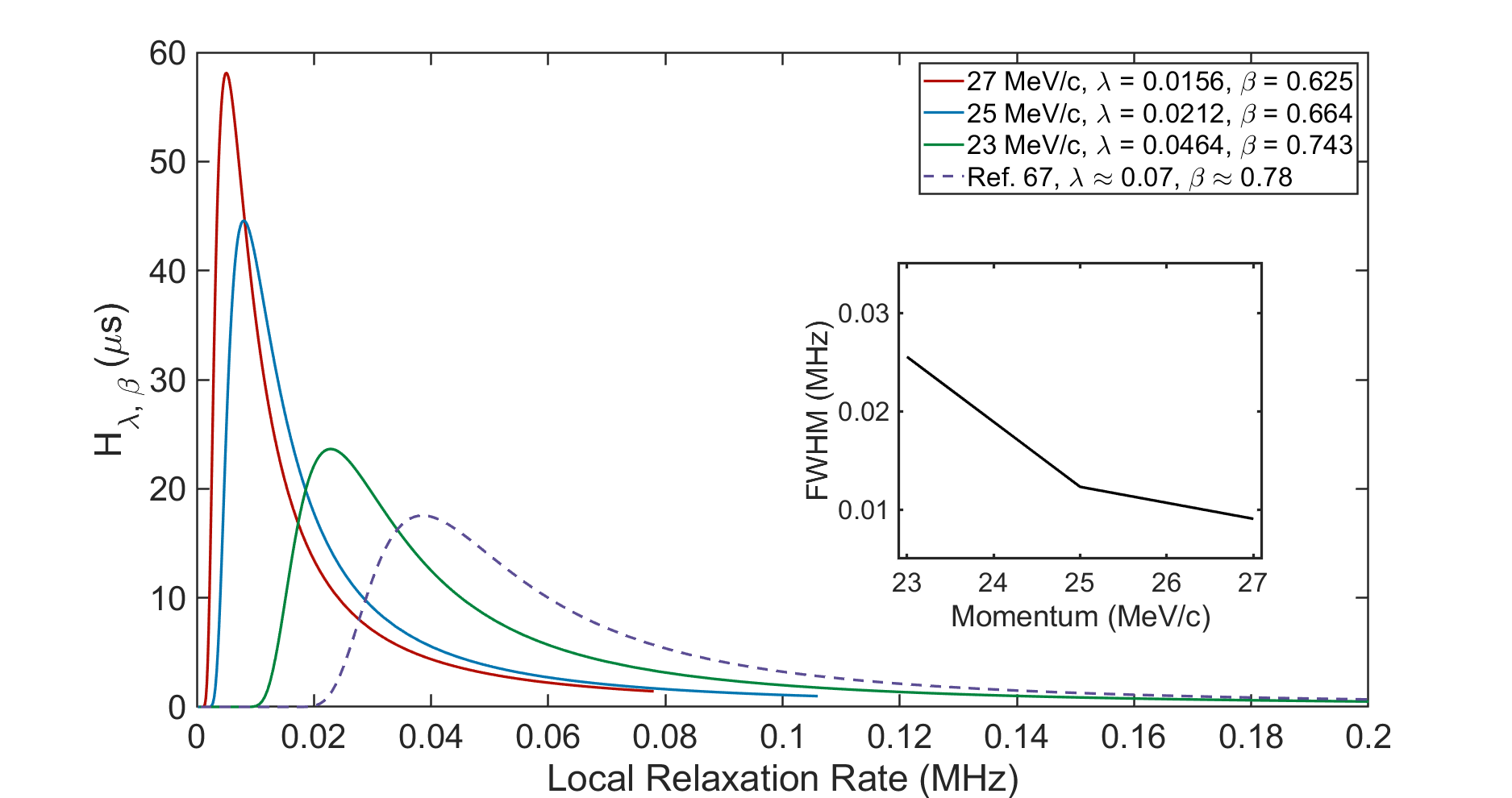}}
	\caption{Distribution of relaxation rates $H_{\lambda, \beta}$ that, when averaged, produce the observed stretched exponential relaxation from the ZF measurements at 5~K. All depths are shown along with the distribution that produces the relaxation observed by Rotaru~\textit{et~al.}~\cite{RotaruPFN}. Inset shows the full width at half maximum (FWHM) against momentum (depth). As the penetration depth increases, the distribution narrows and slows.}
	\label{fig:Dist}
\end{figure*}

Connecting this analysis with our data, \fref{fig:Dist} shows the distribution of local relaxation rates that produce the observed relaxation of the $\mu^+$ polarization under ZF conditions for all momentum transfers. These distributions were computed using~\eref{eq:SeriesForm} as tabulated in~\cite{Dishon}. It shows that the distribution narrows with increasing depth into the sample and that the peak moves left indicating that the relaxation slows (lower frequency) in the bulk. At first this narrowing may seem counter-intuitive as the exponent $\beta$ is usually considered as a measure of the width of the distribution. However, we emphasize that this is only true near $\beta = 1$ where the full width at half maximum (FWHM) increases as $\beta$ decreases from unity for a given value of $\lambda$. The FWHM then encounters a maximum value at $\beta \approx 0.64$ and then gradually decreases to $\beta = 0$.   

This turning point can be further seen by considering the forms of $H_{\lambda, \beta}(s)$ at $\beta = 1$, $1/2$ and 0: For $\beta = 1$, the stretched exponential reduces to a normal exponential decay and so $H_{\lambda, 1}(s) \propto \delta(s-\lambda)$ where $\delta(x)$ is the Dirac delta in order to produce the single relaxation rate. This means that the FWHM must become zero in this limit. At $\beta = 0$, the decay curve is flattened as $e^{-(\lambda t)^0} = e^{-1}$. In this case, $H_{\lambda, 0}(s)$ must also have the form of a Dirac delta but centered at $s=0$ to produce the uniform response. As such, the FWHM must also become zero as $\beta \rightarrow 0$. Finally, as mentioned above, a closed form for $H_{\lambda, 1}(s)$ can be derived for $\beta = 1/2$~\cite{Wagner}. In this case, the FWHM can be calculated as $\approx 0.45 \lambda \neq 0$ and, hence, the frequency FWHM must possess a turning point over the range $0 < \beta < 1$.  

Furthermore, the FWHM will scale linearly with $\lambda$. This dependence may be deduced by considering that $\lambda$ corresponds to a scaling parameter and may be removed by the transforming to a universal distribution \mbox{$G_{\beta}(s) = \lambda H_{\lambda, \beta}(\lambda s)$} (see~\ref{sec:AppendixA}). Critically, the argument above for the $\beta$ dependence of the FWHM also holds for $G_{\beta}(s)$ and so we see that the FWHM of $H_{\lambda, \beta}(\lambda s)$ may be related to that of $G_{\beta}(s)$ by a factor of $\lambda$ corresponding to stretching of the variable $s$.  

This illustrates the importance to plot the frequency distribution for a given parameter $(\beta,  \lambda)$ set over extrapolating results taking the approximation of $\beta=1$.  This is particularly important given the disorder and expected complex local magnetism in PFN.  The depth-narrowing of the FWHM seen in \fref{fig:Dist} is consistent with these two behaviors: $\lambda$ was found to decrease with depth which would be expected to produce a reduction in FWHM with depth. The fitted values of $\beta$ also decreased with depth but, as they are far from unity, this cannot be interpreted as corresponding to broadening. In fact, they are near the turning point at $\beta \approx 0.64$ and so we should expect the FWHM to either remain near stationary or decrease with depth. Thus the combination of these two effects can explain the trend seen in \fref{fig:Dist}.

\begin{table}
\caption{\label{tab:Compare}Table comparing the muon relaxation parameterization for the extreme depths in this study with the literature values taken from Rotaru~\textit{et~al.}~\cite{RotaruPFN}.}
\lineup
\begin{indented}
\item[]\begin{tabular}{@{}ccccc}\br 
	 & & $A_R$ & $\lambda$ (MHz)  & $\beta$ \\\ns \mr
	 \multicolumn{2}{c}{Rotaru~\textit{et~al.} (`Bulk')~\cite{RotaruPFN}} & $\approx$ 1/3 & $\approx$ 0.07 & $\approx$ 0.78 \\ [1ex]
	 \multirow{2}{*}{\rotatebox[origin=cb]{90}{This study}} & Bulk (27 MeV/c) & 1.1(1) & 0.016(1) & 0.63(2) \\ [2ex]
	 & Surface (23 MeV/c) & 0.9(1) & 0.046(2) & 0.74(2) \\ [2ex] \br
\end{tabular}
\end{indented}
\end{table}

\subsection{Comparison to other samples and experiments}

In order to put this into context, the results reported in this paper may be compared to the study performed at PSI by Rotaru~\textit{et~al.}~\cite{RotaruPFN} who also parameterized their decay with a stretched exponential. Two key points are summarized in \tref{tab:Compare} which we now discuss. 

Firstly, the PSI study's relaxing asymmetry begins at $1/3$ below $T_{\textrm{N}}$ whereas the relaxation in this study begins from unity. This can be attributed to the lack of long-range order in this sample as, with the G-type order reported in~\cite{RotaruPFN}, the experiment will only effectively be sensitive to one (out of three) component of the relaxation. No strong evidence of the N\'{e}el transition at $\approx$~140~-~160~K is seen in $\lambda$ or $\beta$ in the sample studied in this report. However, it should be noted that $\lambda$ and $\beta$ are only reported in~\cite{RotaruPFN} for $T < 80$~K so a full comparison of the two cases cannot be made presently in the high temperature limit where the response maybe expected to be similar. 

Secondly, the two PSI relaxation parameters $(\lambda, \beta)$ can be seen to agree more with our lower momentum (`surface') measurements than the higher momentum measurements probing the `bulk'. Further illustrating this, the frequency distribution from the PSI relaxation $(\lambda, \beta)$ parameters is calculated and plotted in \fref{fig:Dist}. This graphically illustrates the similarity between that study and our `surface' measurements and shows that the PSI response contains a much broader range of frequencies.   

\subsection{Skin effect in PFN}

We observe a change in the magnetic relaxation with momentum transfer (and hence implantation depth) over a lengthscale on the order of $\sim$ 100 $\mu$m which is a similar structural lengthscale observed in other relaxors as discussed above.  There is a large change (greater than a factor of 2) in the spin relaxation rate ($\lambda$) with muon momentum over this lengthscale.  This is difficult to reconcile in terms of a small (less than) 10-15\% in Fe$^{3+}$ concentration indicated with negative muon spectroscopy.  We speculate that one possible origin of this skin effect is tied with the a change in lattice constant (or strain) reported using both X-ray~\cite{Kumar20:8} and neutron and X-ray diffraction~\cite{SkinPMN2,Skinreview} in analogous relaxor materials.  Such a mechanism, while over a much larger lengthscale, is similar to suggestions of the differing critical properties reported in the bulk and surface of SrTiO$_{3}$.~\cite{Osterman88:21,McMorrow90:76} We note that our previous neutron spectrosocpy studies of PFN~\cite{StockPFN} found a correlation between the first moment of the magnetic response and also energy scale of the soft optical phonon.  This suggests a link between zone center and finite momentum magnetic dynamics.  We speculate that the strain induced from the formation of ferroelectric order may result in the exchange constants are more frustrated in the near-surface resulting in a broader frequency distribution for the magnetic dynamics as observed here.  It would be interesting to perform analogous studies to that of~\cite{Kong19:29} by oxygen doping, however, given the size of surface areas required for muon spectroscopy such a study would be difficult. We further note that an analogous macroscopic structural skin region has been been observed in thin films of the classical perovskite SrTiO$_{3}$ and tied to changes in strain fields.~\cite{Hunnefeld02:66,Hirota95:52}

\section{Conclusions}\label{Conc}

In conclusion, momentum dependent muon spectroscopy was used to probe the depth dependent properties of a large single crystal sample of the multiferroic PFN. Zero field positive $\mu$SR experiments showed that there is a change in the magnetic properties over this region over the lengthscale of $\sim$ 100 $\mu m$, analogous to the ``skin" region in the structural properties of relaxor-ferroelectrics. By parameterizing the relaxing asymmetry in terms of a phenomenologically motivated stretched exponential, this change may be quantified and compared to literature measurements. Links to compositional heterogeneity were investigated and any possible small changes in the Fe/Nb concentration in the near-surface are difficult to reconcile with the $\mu$SR spectra. Hence, we propose that there is a magnetic skin effect in PFN analogous to the structural skin effect observed in non magnetic relaxor compounds.\nocite{data1,data2}

\ack

The authors would like to thank Ch. Niedermayer for helpful discussions. We are grateful for funding from the EPSRC and the STFC. NGD was supported by EPSRC/Thales UK iCASE Award EP/P510506. 


\appendix

\section{Derivation of an integral form for the stretched exponential distribution}\label{sec:AppendixA}

The stretched exponential function $e^{-(\lambda t)^{\beta}}$ can be motivated as an average over individual exponential relaxations $e^{- s t}$ with the relaxation rate $s$ drawn from a distribution $H_{\lambda, \beta}(s)$~\cite{stretchedExp}:

\begin{equation}
	e^{-(\lambda t)^{\beta}} = \int_{0}^{\infty} ds H_{\lambda, \beta}(s)e^{-st}.
\end{equation}

\noindent The dependence of the effective relaxation rate $\lambda$ can be removed but introducing the variables $T=\lambda t$ and $S = \frac{s}{\lambda}$ so that

\begin{equation}
	e^{-T^{\beta}} = \int_{0}^{\infty} dS G_{\beta}(S)e^{-ST},
\end{equation}

\noindent where the function $G_{\beta}(S)= \lambda H_{\lambda, \beta}(s)$ is the scaled distribution. Clearly, this integral amounts to a Laplace transform~\cite{RHB_13.2} and so $G_{\beta}(S)$ may be calculated by performing the inverse Laplace transform of $e^{-T^{\beta}}$~\cite{BerberanSantos1}.

This result was first obtained by Pollard~\cite{Pollard} but this appendix will outline a method similar to that used by Berberan-Santos \textit{et~al.}~\cite{BerberanSantos1}. These two papers give equivalent integrals and this equivalence will be illustrated here too. A series from for $H_{\lambda, \beta}(s)$ will be derived from this integral in \ref{sec:AppendixB}. 

As standard, the Laplace transform may be inverted by the Bromwich integral

\begin{equation}\label{eq:bromInv}
	G_{\beta}(S) = \lim_{\gamma \rightarrow 0} \frac{1}{2 \pi i} \int_{\gamma - i \infty}^{\gamma + i \infty} dT e^{-T^{\beta} + ST},
\end{equation}

\noindent where the customary $\gamma$ allows the integration contour to be shifted away from $T=0$ where the integrand has a branch point. Taking the branch cut along the negative real axis, the integration variable may be split into real and imaginary components $T = \gamma + iy$. Furthermore, a polar representation of $T$ may also be implemented where $T = Re^{i\theta}$ and this will be used in the following to show that the imaginary part of the integral vanishes due to symmetry. Combining these representations give $R = \frac{\gamma}{\cos \theta}$ and $T = \frac{\gamma e^{i \theta}}{\cos \theta} = \gamma \big[ 1+ i \tan \theta \big]$. Substitution into~\eref{eq:bromInv} then gives an integral purely in terms of the angle $\theta$ as shown in~\eref{eq:AppInt1}. After some manipulation using trigonometric identities, this may be rewritten as~\eref{eq:AppInt2} where $\phi =\Big( \frac{\gamma}{\cos \theta} \Big) ^{\beta} \sin(\beta \theta) - S \gamma \tan \theta$.

\begin{equation}\label{eq:AppInt1}
\eqalign{G_{\beta}(S) = & \lim_{\gamma \rightarrow 0} \frac{1}{2 \pi} \int_{- \frac{\pi}{2}}^{\frac{\pi}{2}} d\theta \frac{\gamma}{\cos^2 \theta} \exp\Big\{-\Big( \frac{\gamma}{\cos \theta} \Big) ^{\beta}\big[\cos(\beta \theta) + i \sin(\beta \theta) \big] + \\
& S \gamma (1 + i \tan \theta) \Big\}.}
\end{equation}

\begin{equation}\label{eq:AppInt2}
\eqalign{G_{\beta}(S) = & \lim_{\gamma \rightarrow 0} \frac{1}{2 \pi} \int_{- \frac{\pi}{2}}^{\frac{\pi}{2}} d\theta \frac{\gamma}{\cos^2 \theta} \exp\Big\{-\Big( \frac{\gamma}{\cos \theta} \Big) ^{\beta}\cos(\beta \theta) + S \gamma \Big\}\\
&\times \big [ \cos \phi - i \sin \phi \big ].}
\end{equation}

Now, by observing that $\phi$ is an odd function of $\theta$, together with the fact that the other terms in the integrand are even, means that the real part is totally even and the imaginary part is totally odd. Hence, since the limits of the integral are symmetric, the odd part must vanish due to symmetry. This leaves

\begin{equation}
	G_{\beta}(S) = \lim_{\gamma \rightarrow 0} \frac{1}{\pi} \int_{0}^{\frac{\pi}{2}} d\theta  \frac{\gamma}{\cos^2 \theta}\exp\Big\{-\Big( \frac{\gamma}{\cos \theta} \Big) ^{\beta}\cos(\beta \theta) + S \gamma \Big\} \cos \phi,
\end{equation}

\noindent where the symmetry has also been used to reduce the integration range to positive angles. 

Having accomplished the goal of removing the imaginary part, let us now reintroduce the variable $y = \gamma \tan \theta$ and transform the integral to

\begin{equation}\label{eq:Appnearlythere}
\eqalign{G_{\beta}(S) = & \lim_{\gamma \rightarrow 0} \frac{1}{\pi} \int_{0}^{\infty} dy \exp\Big\{S \gamma - (\gamma^2 + y^2) ^{\frac{\beta}{2}}\cos\big(\beta \theta(y)\big) \Big\} \\
&\times \cos \Big [(\gamma^2 + y^2) ^{\frac{\beta}{2}}\sin\big(\beta \theta(y)\big) - Sy \Big ].}
\end{equation}

The final steps in order to arrive at the Berberan-Santos form for $H_{\lambda, \beta}(s)$ is to impose the limit $\gamma \rightarrow 0$ and counteract the scaling the distribution. As $y = \gamma \tan \theta$, in order for $y$ to remain finite as $\gamma \rightarrow 0$, we should set $\theta \rightarrow \frac{\pi}{2}$ so that the tangent diverges. Hence, in order to fully account for this limit we set $\gamma \rightarrow 0$ \emph{and} $\theta \rightarrow \frac{\pi}{2}$ in~\eref{eq:Appnearlythere} to give

\begin{equation}
	G_{\beta}(S) = \frac{1}{\pi} \int_{0}^{\infty} dy \exp\Big\{-y^{\beta}\cos\big(\frac{\beta\pi}{2}\big)\Big\} \cos \Big [y^{\beta}\sin\Big(\frac{\beta\pi}{2}\Big) - Sy \Big ].
\end{equation}

\noindent Then by removing the scaling we arrive at an integral form for the distribution of relaxation rates

\begin{equation}\label{eq:AppBSForm}
	H_{\lambda, \beta}(s) = \frac{1}{\pi \lambda} \int_{0}^{\infty} dy \exp\Big\{-y^{\beta}\cos\big(\frac{\beta\pi}{2}\big)\Big\} \cos \Big [y^{\beta}\sin\Big(\frac{\beta\pi}{2}\Big) - \frac{sy}{\lambda} \Big ].
\end{equation}

However, Pollard first derived an alternative integral form, again using contour integration~\cite{Pollard} 

\begin{equation}\label{eq:AppPollardForm}
	H_{\lambda, \beta}(s) = \frac{1}{\pi \lambda} \int_{0}^{\infty} du \exp\Big\{-\frac{su}{\lambda}-u^{\beta}\cos(\beta\pi)\Big\} \sin \Big [u^{\beta}\sin(\beta\pi)\Big ].
\end{equation}

\noindent We will now show the equivalence of~\eref{eq:AppBSForm} and~\eref{eq:AppPollardForm}.

First observe that~\eref{eq:AppBSForm} maybe be rewritten as

\begin{equation}
\eqalign{H_{\lambda, \beta}(s) & = \mathcal{R}e\Bigg\{ \frac{1}{\pi \lambda} \int_{0}^{\infty} dy \exp\Big\{-y^{\beta}\cos\big(\frac{\beta\pi}{2}\big)+i \big [y^{\beta}\sin\big(\frac{\beta\pi}{2}\big) - \frac{sy}{\lambda} \big ]\Big\} \Bigg\} \\
& = \mathcal{R}e\Bigg\{ \frac{1}{\pi \lambda} \int_{0}^{\infty} dy \exp\Big\{-\big(y e^{\frac{-i\pi}{2}}\big)^{\beta} - \frac{isy}{\lambda}\Big\} \Bigg\}.}
\end{equation}

\noindent As we have already shown that the imaginary part of this is zero, this more compact representation is valid. Performing a rotation in the complex plane, let $u = iy$ to give

\begin{equation}\label{eq:Compact}
\eqalign{H_{\lambda, \beta}(s) & = \mathcal{R}e\Bigg\{ \frac{-i}{\pi \lambda} \int_{0}^{\infty} du \exp\Big\{-\big(u e^{-i\pi}\big)^{\beta} - \frac{su}{\lambda}\Big\} \Bigg\} \\
	& = \mathcal{I}m\Bigg\{ \frac{1}{\pi \lambda} \int_{0}^{\infty} du \exp\Big\{-\big(u e^{-i\pi}\big)^{\beta}- \frac{su}{\lambda}\Big\} \Bigg\}.}
\end{equation}

\noindent Finally by expanding the complex exponentials, we arrive at

\begin{equation}
\eqalign{H_{\lambda, \beta}(s) & = \mathcal{I}m\Bigg\{ \frac{1}{\pi \lambda} \int_{0}^{\infty} du \exp\Big\{-u^{\beta}\cos(\beta \pi) + i u^{\beta}\sin(\beta \pi) - \frac{su}{\lambda}\Big\} \Bigg\} \\
	& = \frac{1}{\pi \lambda} \int_{0}^{\infty} du \exp\Big\{- \frac{su}{\lambda} - u^{\beta}\cos(\beta \pi)\Big\} \mathcal{I}m\Big\{ e^{i u^{\beta}\sin(\beta \pi)}\Big\},}
\end{equation}

\noindent which, when the imaginary part is taken, gives~\eref{eq:AppPollardForm}. As noted by Berberan-Santos~\textit{et~al.}, these two integral representations are actually complementary. Equation~\ref{eq:AppPollardForm} is easier to compute numerically at large $s$ than~\eref{eq:AppBSForm} due to the large oscillations of the integrand. The opposite is true for small $s$~\cite{BerberanSantos1}.

\section{Further derivation of a series form for the stretched exponential distribution}\label{sec:AppendixB}

Returning to~\eref{eq:Compact}, we may alternatively expand the first exponential

\begin{equation}
	H_{\lambda, \beta}(s) = \mathcal{I}m\Bigg\{ \frac{1}{\pi \lambda} \int_{0}^{\infty} du \sum_{n=0}^{\infty}\frac{1}{n!} \Big [-u^{\beta} e^{-i\beta \pi} \Big ]^n e^{- \frac{su}{\lambda}} \Bigg\}.
\end{equation}

Integrating term by term (permitted since the power series in convergent) this gives

\begin{equation}
	H_{\lambda, \beta}(s) = \mathcal{I}m\Bigg\{ \frac{1}{\pi \lambda}  \sum_{n=0}^{\infty}\frac{e^{-i\beta n \pi}}{n!} (-1)^n \int_{0}^{\infty} du u^{\beta n} e^{- \frac{su}{\lambda}} \Bigg\},
\end{equation}

\noindent which, along with the substitution $x = \frac{su}{\lambda}$ allows us to recognize the gamma function

\begin{equation}
\eqalign{H_{\lambda, \beta}(s) & = \mathcal{I}m\Bigg\{ \frac{1}{\pi \lambda}  \sum_{n=0}^{\infty}\frac{e^{-i\beta n \pi}}{n!} (-1)^n \Big(\frac{\lambda}{s} \Big)^{\beta n +1} \int_{0}^{\infty} dx x^{\beta n} e^{- x} \Bigg\} \\
	 & = \mathcal{I}m\Bigg\{ \frac{1}{\pi \lambda}  \sum_{n=0}^{\infty}\frac{e^{-i\beta n \pi}}{n!} (-1)^n \Big(\frac{\lambda}{s} \Big)^{\beta n +1} \Gamma (1 + \beta n) \Bigg\}.}
\end{equation}

Lastly, we may observe that, apart from the complex exponential, the whole expression is real and so taking the imaginary part simply transforms this exponential to a sine function. Then, the odd property of sine may be used along with the fact that $\sin 0 = 0$ (meaning that $n=0$ term correspondingly vanishes) to give the series form for the distribution $H_{\lambda, \beta}(s)$

\begin{equation}
	H_{\lambda, \beta}(s) = \frac{1}{\pi \lambda}  \sum_{n=1}^{\infty}\frac{\sin(\beta n \pi)}{n!} (-1)^{n+1} \Big(\frac{\lambda}{s} \Big)^{\beta n +1} \Gamma (1 + \beta n). \\
\end{equation}

\noindent This series was first derived by Humbert and is convergent for all $s$~\cite{Humbert}. It is this form that was used by Dishon~\textit{et~al.} to compute $H_{\lambda, \beta}(s)$~\cite{Dishon}, the values of which were used in this work.


\end{document}